\documentclass[twocolumn]{autart} %\documentclass[journal,twoside,web]{ieeecolor}

\usepackage{amsmath,amssymb,amsfonts}
\usepackage{algorithmic}
\usepackage{graphicx}
\usepackage{algorithm,algorithmic}

\usepackage{textcomp}
\usepackage{float}
\usepackage{mathtools}
\usepackage{bm}

\usepackage{mathtools}
\usepackage{xurl}
\usepackage{mathrsfs}
\usepackage{xcolor}
\usepackage{float}
\usepackage{afterpage}
\usepackage{etoolbox} % for \ifstrempty

\newcounter{kinfcounter}
\renewcommand{\thekinfcounter}{\arabic{kinfcounter}}

\newcommand{\kfunc}[2][]{%
  \refstepcounter{kinfcounter}%
  \expandafter\gdef\csname alpha\thekinfcounter\endcsname{#2}%
  \ifstrempty{#1}{}{%
    \label{#1}%
  }%
  \[
    \alpha_{\thekinfcounter}(r) = #2
  \]
}

\newcommand{\reusekfunc}[1]{%
  \[
    \alpha_{#1}(r) = \csname alpha#1\endcsname
  \]
}

\newcommand{\qedsymbol}{$\blacksquare$}
\usepackage{etoolbox}
\AtEndEnvironment{pf}{\hfill\qedsymbol}

\newtheorem{definition}{Definition}
\newtheorem{theorem}{Theorem}
\newtheorem{lemma}{Lemma}
\newtheorem{assumption}{Assumption}
\newcommand{\todo}[1]{\color{red}#1\color{black}}

\newlength{\leftstackrelawd}
\newlength{\leftstackrelbwd}
\def\leftstackrel#1#2{\settowidth{\leftstackrelawd}%
{${{}^{#1}}$}\settowidth{\leftstackrelbwd}{$#2$}%
\addtolength{\leftstackrelawd}{-\leftstackrelbwd}%
\leavevmode\ifthenelse{\lengthtest{\leftstackrelawd>0pt}}%
{\kern-.5\leftstackrelawd}{}\mathrel{\mathop{#2}\limits^{#1}}}

\newcommand{\sgn}{\text{sgn}}

\begin{document}
\begin{frontmatter}
\title{Stabilization of nonlinear systems with unknown delays via \\  delay-adaptive neural operator approximate predictors}
\author%[author1]
{Luke Bhan}$^\text{a,}$\ead{lbhan@ucsd.edu}$^\ast$\corauth{Corresponding author},\thanks{\textit{Acknowledgements:} The work of L. Bhan is supported by the U.S. Department of Energy (DOE) grant DE-SC0024386. The work of M. Krstic was funded by AFOSR grant FA9550-23-1-0535 and NSF grant ECCS-2151525. The work of Y. Shi is supported by Department of Energy grant DE-SC0025495, Schmidt Sciences AI2050 Early Career Fellowship, and NSF grant ECCS-2442689. }   \
\author%[author2]
{Miroslav Krstic}$^\text{a}$\ead{krstic@ucsd.edu},\ 
\author
{Yuanyuan Shi}$^\text{a}$\ead{yyshi@ucsd.edu} 
\address {$^\text{a}$ University of California, San Diego, USA}

\begin{abstract}
 This work establishes the first rigorous stability guarantees for approximate predictors in delay-adaptive control of nonlinear systems, addressing a key challenge in practical implementations where exact predictors are unavailable. We analyze two scenarios: (i) when the actuated input is directly measurable, and (ii) when it is estimated online. For the measurable input case, we prove semi-global practical asymptotic stability with an explicit bound proportional to the approximation error $\epsilon$. For the unmeasured input case, we demonstrate local practical asymptotic stability, with the region of attraction explicitly dependent on both the initial delay estimate and the predictor approximation error. To bridge theory and practice, we show that neural operators—a flexible class of neural network-based approximators—can achieve arbitrarily small approximation errors, thus satisfying the conditions of our stability theorems. Numerical experiments on two nonlinear benchmark systems—a biological protein activator/repressor model and a micro-organism growth Chemostat model—validate our theoretical results. In particular, our numerical simulations confirm stability under approximate predictors, highlight the strong generalization capabilities of neural operators, and demonstrate a substantial computational speedup of up to $15\times$ compared to a baseline fixed-point method.
 \end{abstract}

\end{frontmatter}
\allowdisplaybreaks
\setlength{\parskip}{.5em} 

%
% Start line numbering here if you want
%

% I think I should rewrite this introduction to be more focused on the results of the first approximate predictors for delay adaptive control. 

\section{Introduction}
In this work, we extend the concept of approximate predictors using operator learning to delay systems with constant but \emph{unknown} delays. The presence of unknown delays presents a significant challenge, as the approximation of the predictor introduces an additive error requiring analysis akin to robust adaptive control. Furthermore, unlike robust adaptive control, the adaptive parameter estimated is the actuator delay which directly impacts system stability. 
Despite these challenges, we achieve results similar to \cite{pmlr-v283-bhan25a}, ensuring the semi-global practical stability of the feedback system, which depends on the approximation error and an additional error term on the delay bounds. This represents the first result for implementing any type of approximate predictor when the delay is constant but unknown.

\subsection{Predictor feedback designs and implementations} \label{subsec:introduction-predictor-feedback}
Predictor feedback methods for compensating actuator delays in dynamical systems have been studied for over 50 years \cite{smith1957closer,henson1994time,690350914,935057,1272267,1272269,doi:10.1137/040616383}. These approaches have proven effective across diverse applications such as traffic control \cite{9987680,6248674}, aerospace vehicles \cite{8930010,9468382}, and robotics \cite{9727201,BEKIARISLIBERIS20131576,7458842}. However, in nonlinear systems, implementation remains challenging because the predictor is defined by an implicit ordinary differential equation \cite{nikoBook}. To address this, \cite{iassonBook} proposed numerical schemes combining finite differencing with successive approximations to approximate predictors for both linear and nonlinear systems, along with stability guarantees. Yet, this work does not consider delay-adaptive cases, leaving stability for predictors with unknown delays unestablished. Moreover, these schemes are computationally expensive due to the need for fine discretization. Recently, \cite{pmlr-v283-bhan25a} introduced operator learning to approximate the predictor mapping, achieving speedups of 1000× compared to traditional solvers, but this was limited to nonlinear systems with constant, known delays. In this work, we extend \cite{pmlr-v283-bhan25a} to handle unknown delays, addressing a critical gap in delay-adaptive predictor feedback.

While exact predictor feedback for unknown delays was initially addressed in \cite{delphine}, introducing an approximate predictor significantly complicates the analysis. Now, additive errors arise in the Lyapunov bounds, which still permit an ISS-like stability result when the distributed input is known. However, when both the actuator input and delay are unknown, additional multiplicative approximation errors emerge, restricting stability to a smaller, local region of attraction depending both on the neural approximation error and the initial delay estimation error. This is similar to the results of robust adaptive control \cite{486648,4793094}- although significantly more challenging to the adaptive parameter being the delay on the actuator. Nonetheless, we show that this region persists as long as the approximation errors remain sufficiently small \textbf{and} the initial delay estimation error is close enough. 

\subsection{Operator learning in control} \label{subsec:operator-learning}
We briefly discuss the advances of operator learning as a tool for real-time implementation of controllers. Neural operators were first introduced in \cite{392253} and then more recently popularized for learning PDE solutions in \cite{li2021fourier} and \cite{Lu2021}. The goal is to approximate any infinite dimensional operator between function spaces by finite dimensional neural networks. In control theory, operator learning was first introduced in \cite{bhan2023neural,KRSTIC2024111649} for approximating the gain kernel in linear PDEs, and has now expanded to output-feedback PDEs \cite{XIAO2024106620}, delayed PDEs systems \cite{10872816,QI2024105714}, gain-scheduling \cite{10918744}, adaptive control of PDEs \cite{lamarque2024adaptiveneuraloperatorbacksteppingcontrol,bhan2024adaptivecontrolreactiondiffusionpdes} and predictor feedback designs \cite{pmlr-v283-bhan25a}.
Notably, \cite{pmlr-v242-zhang24c} presented the first application of operator learning in real-world traffic scenarios, marking a significant milestone for practical deployment. We further review neural operators in Section \ref{sec:bg-neural-operators}.

\subsection{Notation}
For functions, $h: [0, D] \times \mathbb{R} \to \mathbb{R}$, we use $h_x(x, t) = \frac{\partial h}{\partial x}(x, t)$ and $h_t(x, t) = \frac{\partial h}{\partial t}(x, t)$ to denote derivatives. We use $C^1([t-D, t];\mathbb{R}^m)$ to denote the set of functions with continuous first derivatives mapping the interval $[t-D, t]$ to $\mathbb{R}^m$. For a $n$-vector, we use $|\cdot|$ for the Euclidean norm. For functions, we define the spatial $L^p$ norms as $\|h(t)\|_{L^p[0, D]} = (\int_0^D |h(x, t)|^p dx)^{\frac{1}{p}}$ for $p\in [1, \infty)$. For brevity, we use the shorthand \(\alpha_{i:j} \in \mathcal{K}_\infty\) to indicate that the functions \(\alpha_i, \alpha_{i+1}, \ldots, \alpha_j\) are all elements of \(\mathcal{K}_\infty\).

\color{blue} A preliminary version appeared at CDC 2025 as "Delay-adaptive Control of Nonlinear Systems with Approximate Neural Operator Predictors." This journal version adds new theory for unmeasured actuator input (Section~\ref{sec:unmeasured}) and new Chemostat simulations (Section~\ref{sec:chemostat}).

\color{black}

\section{Technical background} \label{sec:background}
\subsection{Delay-adaptive control} \label{sec:bg-delay-adpative}
We study the nonlinear plant
\begin{align}
    \dot{X}(t) &= f(X(t), U(t-D))\,, \label{eq:main-problem}
\end{align}
where $X \in \mathbb{R}^n$, $U \in \mathbb{R}$, $f \in C^2(\mathbb{R}^n \times \mathbb{R}; \mathbb{R}^n)$ such that $f(0, 0) = 0$, and $D$ is an unknown delay within the interval $[\underline{D}, \overline{D}]$ where $\underline{D} > 0$. As standard in the predictor feedback literature \cite{delphine}, we introduce the following assumptions:
\begin{assumption} \label{assumption:strongly-forward-complete}
    The plant $\dot{X} = f(X, \Omega)$ with $\Omega$ scalar is strongly forward complete. 
\end{assumption}
\begin{assumption} \label{assumption:gas}
    There exists $\kappa \in C^2(\mathbb{R}^n; \mathbb{R})$ such that the feedback law  $U(t) = \kappa(X(t))$ guarantees that the delay-free plant is globally exponentially stable. 
\end{assumption}

Note that, under Assumption \ref{assumption:gas}, \cite[Theorem 4.14]{khalil} implies that there exists $\lambda > 0$ and a class $C^\infty$ radially unbounded positive definite function $V$ such that for any $X \in \mathbb{R}$, we have 
\begin{align}
    \frac{dV}{dX}(X) f(X, \kappa(X)) &\leq -\lambda V(X) \,, \label{eq:linear-lyapunov-1} \\
    |X|^2 &\leq V(X) \leq C_1 |X|^2\,,\label{eq:linear-lyapunov-2} \\ 
    \left|\frac{dV}{dX}(X) \right| &\leq C_2 |X| \label{eq:linear-lyapunov-3} \,, 
\end{align}
where $\lambda, C_1, C_2 > 0$. 

\begin{assumption}\label{assumption:lipschitz-dynamics}
Let $f(X, U)$ be the plant dynamics as in \eqref{eq:main-problem} and $\mathcal{X} \subset \mathbb{R}^n$ and $\mathcal{U} \subset \mathbb{R}$ be compact domains with bounds $\overline{X}$, $\overline{U}$ respectively. Then, there exists a constant $C_f(\overline{X}, \overline{U}) > 0$ such that $f$ satisfies the Lipschitz condition
\begin{align} \label{eq:lipschitz-dynamics-cond}
    |f(X_1, u_1) - f(X_2, u_2)| \leq C_f(|X_1-X_2| + |u_1-u_2|)\,,
\end{align}
for all $X_1, X_2 \in \mathcal{X}$ and $u_1, u_2 \in \mathcal{U}$. 
\end{assumption}

\begin{assumption} \label{assumption:growth-condition}
    The function $f$, the Jacobian matrix $\partial f/\partial X$, the control law $\kappa$ and its derivative satisfy the following growth conditions
    \begin{alignat}{2}
        |f(X, U)| &\leq M_1(|X|+|U|)\,,& \quad \left| \frac{\partial f}{\partial X}(X, U) \right| &\leq M_2 \,, \nonumber  \\
        |\kappa(X)| &\leq M_3|X|\,,& \quad \left| \frac{d\kappa}{dX}(X) \right| &\leq M_4\,,\nonumber 
    \end{alignat}
    for constants $M_1, M_2, M_3, M_4 > 0$.
\end{assumption}

Assumption \ref{assumption:strongly-forward-complete} ensures the plant \eqref{eq:main-problem} does not escape before the input reaches the system at $t=D$. Assumption \ref{assumption:gas} of global exponential stability is needed for the semi-global results in Section \ref{sec:theoretical-analysis}, though it can be relaxed for a local result. Assumption \ref{assumption:lipschitz-dynamics}- Lipschitz dynamics on a compact set - is required for establishing the existence of an approximate predictor. Assumption \ref{assumption:growth-condition} is needed to show global asymptotic stability without any predictor approximation. Assumption \ref{assumption:growth-condition} makes Assumption \ref{assumption:strongly-forward-complete} redundant and is partially implied by Lipschitzness in Assumption \ref{assumption:lipschitz-dynamics}, but is included for clarity and consistency with the exact predictor feedback design \cite{delphine}.

As standard in predictor feedback designs, we reformulate the plant \eqref{eq:main-problem} into the following ODE-PDE cascade where the delay is absorbed into the transport PDE with velocity $D$:
\begin{subequations}
\begin{align}
    \dot{X}(t) &= f(X(t), u(0, t))\,, \label{eq:plant-ode-pde-1} \\
    Du_t(x, t) &= u_x(x, t) \,, \label{eq:plant-ode-pde-2}\\ 
    u(1, t) &= U(t)\,.  \label{eq:plant-ode-pde-3}
\end{align}
\end{subequations}

The representation of the delay through a transport PDE was first introduced in \cite{miroslavDelay} where the analytical solution of the transport PDE recovers the original plant. Namely, we have that
\begin{align} \label{eq:transport-pde-sol}
    u(x, t) = U(t+D(x-1))\,, \quad  x \in [0, 1]\,, 
\end{align}
and thus when $x=0$ in the PDE representation, we recover $U(t-D)$ as in \eqref{eq:main-problem}. To compensate for the delay, we estimate the state $D$ timesteps in the future with the predictor given for all $x\in [0, 1]$ 
\begin{align} \label{eq:exact-predictor-no-adaptive}
    p(x, t) = X(t+Dx) = X(t) + D \int_0^x f(p(y, t), u(y, t))dy\,, 
\end{align}
which is itself, an implicit ODE whose analytical form is unknown for most nonlinear plant dynamics $f$. Then, under the assumption that the delay is known, the stabilizing control law is given by $U(t) = \kappa(p(1, t))$. When the delay is unknown, as in the problem setting of this work, the control law is chosen by 
\begin{align} \label{eq:exact-controller-adaptive}
    U(t) = \kappa(\breve{p}(1, t))\,,
\end{align}
where the predictor is given by
\begin{align} \label{eq:exact-predictor-adaptive}
    \breve{p}(x, t) = X(t) + \breve{D}(t) \int_0^x f(\breve{p}(y, t), u(y, t)) dy\,. 
\end{align}
with the delay estimate $\breve{D}(t)$. 
Following \cite{delphine}, $\breve{D}(t)$ is then updated via 
\begin{align} \label{eq:update-law-projector}
    \dot{\breve{D}}(t) &= \gamma \text{Proj}_{[\underline{D}, \overline{D}]} \left\{ \breve{D}(t), \phi(t) \right\}\,, \\ \label{eq:update-law}
    \phi(t) &= - \frac{\int_0^1 (1+x) q_1(x, t)w(x, t)dx}{1 + V(x) + b\int_0^1(1+x)w(x, t)^2dx}\,, 
\end{align}
where $b$ is a user-specified parameter, $V$ is the positive definite Lyapunov function in \eqref{eq:linear-lyapunov-1}, \eqref{eq:linear-lyapunov-2}, \eqref{eq:linear-lyapunov-3}, $w$ is given by the backstepping transformation
\begin{align} \label{eq:w-def-bcks-transform}
    w(x, t) = u(x, t) - \kappa(\breve{p}(x, t))\,, 
\end{align}
and the scalar function $q_1$ is defined as 
\begin{align}
    q_1(x, t) = \frac{d \kappa}{d \breve{p}}(\breve{p}(x, t)) \Phi(x, 0, t) f(\breve{p}(0, t), u(0, t))\,, 
\end{align}
where $\Phi$ is the transition matrix associated with the space-varying time-parameterized equation \newline 
$(dr/dx)(x) = \breve{D}(t) (\partial f/\partial \breve{p}) (\breve{p}(x, t), u(x, t))r(x)$.

Then, under the control feedback law \eqref{eq:exact-controller-adaptive}, \cite[Theorem 1]{delphine} states that the plant is globally asymptotically stable. However, as mentioned, \eqref{eq:exact-predictor-adaptive} is not analytically known and therefore needs to be approximated for any real-world implementation. Thus, we introduce an approximation for the predictor $\breve{p}$ and show that, under the universal approximation of neural operators, semi-global practical asymptotic stability is achieved. 

\subsection{Neural operators} \label{sec:bg-neural-operators}
For the readers reference, we briefly review the related background and corresponding theoretical results used in the analysis in Section \ref{sec:theoretical-analysis}. Neural operators are finite dimensional approximations of nonlinear operators (e.g. the solution operator to any nonlinear ODE) which map across function spaces. Thus, a neural operator takes a representation of the input function $c(x)$ and its evaluation point $x$ as inputs. Additionally, as input, it takes a value $y$, which specifies the point to evaluate the target function after applying the operator to $c$. The neural operator then provides an approximation of the value of the operator applied to $c(x)$ at the point $y$ in the target function space via finite neural networks.

We now describe neural operators from a rigorous perspective. Namely, let $\Omega_u \subset \mathbb{R}^{d_{u_1}}$, $\Omega_v \subset \mathbb{R}^{d_{v_1}}$ be bounded domains and let  $\mathcal{F}_c \subset C^0(\Omega_u; \mathbb{R}^c)$, $\mathcal{F}_v \subset C^0(\Omega_v; \mathbb{R}^v)$ be continuous function spaces. Then, we define a neural operator approximation for the nonlinear operator $\Psi$ as any  function satisfying the following form:
\begin{definition} \label{definition:neural-operator} \cite[Section 1.2]{lanthaler2024nonlocalitynonlinearityimpliesuniversality}
    Given the parameter for channel dimension $d_c$, we call any $\hat{\Psi}$ a neural operator given it satisfies the compositional form $\hat{\Psi} = \mathcal{Q} \circ \mathcal{L}_L \circ \cdots \circ \mathcal{L}_1 \circ \mathcal{R}$ where  $\mathcal{R}$ is a lifting layer, $\mathcal{L}_l, l=1,..., L$ are the hidden layers, and $\mathcal{Q}$ is a projection layer. That is, 
    $\mathcal{R}$ is given by 
    \begin{equation}
    \mathcal{R} : \mathcal{F}_c(\Omega_u; \mathbb{R}^c) \rightarrow \mathcal{F}_s(\Omega_s; \mathbb{R}^{d_c}), \quad c(x) \mapsto R(c(x), x)\,, 
\end{equation} where $\Omega_s \subset \mathbb{R}^{d_{s_1}}$, $\mathcal{F}_s(\Omega_s; \mathbb{R}^{d_c})$ is a Banach space for the hidden layers and $R: \mathbb{R}^c \times \Omega_u \rightarrow \mathbb{R}^{d_c}$ is a learnable neural network acting between finite-dimensional Euclidean spaces. For $l=1, ..., L$, each hidden layer is given by 
\begin{equation} \label{eq:generalNeuralOperator}
    (\mathcal{L}_l v)(x) := s \left( W_l v(x) + b_l + (\mathcal{K}_lv)(x)\right)\,, 
\end{equation}
where weights $W_l \in \mathbb{R}^{d_c \times d_c}$ and biases $b_l \in \mathbb{R}^{d_c}$ are learnable parameters, $s: \mathbb{R} \rightarrow \mathbb{R}$ is a smooth, infinitely differentiable activation function that acts component wise on inputs and $\mathcal{K}_l$ is the nonlocal operator given by 
\begin{equation} \label{eq:generalKernel}
    (\mathcal{K}_lv)(x) = \int_\mathcal{X} K_l(x, y) v(y) dy\,,
\end{equation}
where $K_l(x, y)$ is a kernel function containing learnable parameters. Lastly, the projection layer $\mathcal{Q}$ is given by 
\begin{equation}
    \mathcal{Q} : \mathcal{F}_s(\Omega_s; \mathbb{R}^{d_c}) \rightarrow \mathcal{F}_v(\Omega_v; \mathbb{R}^v), \quad s(x) \mapsto Q(s(x), y)\,, 
\end{equation}
where $Q$ is a finite dimensional neural network from $\mathbb{R}^{d_c} \times \Omega_v \rightarrow \mathbb{R}^v$. 
\end{definition}

This abstract formulation of neural operators covers a wide range of architectures including the Fourier Neural Operator (FNO) \cite{li2021fourier} as well as DeepONet \cite{Lu2021} where the differentiator between these approaches lies in the implementation of the kernel function $K_l$. However, in \cite{lanthaler2024nonlocalitynonlinearityimpliesuniversality}, the authors showed that a single hidden layer neural operator with a kernel given by the averaging kernel $K_l(x, y) =1/|\mathcal{X}|$ where $|\mathcal{X}|$ is the diameter of the domain is sufficient for universal approximation in the following sense
\begin{theorem}
    \label{thm:neural-operator-uat}
    \cite[Theorem 2.1]{lanthaler2024nonlocalitynonlinearityimpliesuniversality} Let $\Omega_u \subset \mathbb{R}^{d_{u_1}}$ and $\Omega_v \subset \mathbb{R}^{d_{v_1}}$ be two bounded domains with Lipschitz boundary. Let $\Psi: C^0(\overline{\Omega_u};\mathbb{R}^{d_{u_1}}) \rightarrow C^0(\overline{\Omega_v}; \mathbb{R}^{d_{v_1}}) $ be a continuous operator and fix a compact set $K \subset C^0(\overline{\Omega_u};\mathbb{R}^{d_{u_1}})$. Then for any $\epsilon > 0$, there exists a channel dimension $d_c > 0$ such that a single hidden layer neural operator $\hat{\Psi}: K \rightarrow C^0(\overline{\Omega_v}; \mathbb{R}^{d_{v_1}})$ with kernel function $K_l(x, y) =1/|\mathcal{X}|$ satisfies  
\begin{equation}
    \sup_{u \in K} |\Psi(u)(y) - \hat{\Psi}(u)(y)|\leq \epsilon\,,
\end{equation}
for all values $y \in \Omega_v$.
\end{theorem}

The key corollary to Theorem 1 is that, many of the architectures aforementioned (e.g. FNO, Laplace neural operator, DeepONet) contain kernel functions $K_l$ that recover the averaging kernel and thus are universal under the setting of Theorem \ref{thm:neural-operator-uat}.  
Henceforth, for the remainder of this study, we assume that any reference to a neural operator is an architecture of the form given in Definition \ref{definition:neural-operator} with a viable kernel function such that Theorem \ref{thm:neural-operator-uat} holds. We are now ready to introduce the predictor operator to be approximated. 
\section{Neural operator approximate predictors} \label{sec:theoretical-analysis}
\begin{definition} \label{defintion:predictor-operator}
    Let $X \in \mathbb{R}^n$, $U \in C^2([0, 1]; \mathbb{R})$, $\varphi \in \mathbb{R}^+$. Then, we define the \textbf{predictor operator} as the mapping $\breve{\mathcal{P}}: (X, U, \varphi) \rightarrow \breve{P}$ where $\breve{P}(s) = \breve{\mathcal{P}}(X, U, \varphi)(s)$ satisfies for all $s \in [0, 1]$, 
    \begin{align} \label{eq:predictor-operator}
        \breve{P}(s) - X - \varphi \int_0^s f(\breve{P}(s), U(s)) ds = 0\,. 
    \end{align}
\end{definition}

Notice that, by definition the predictor operator yields the solution to \eqref{eq:exact-predictor-adaptive} where $\varphi$, as an input to the operator, is the estimated delay. Furthermore, in order to approximate this mapping with a neural operator such that Theorem \ref{thm:neural-operator-uat} holds, we require that the predictor operator is continuous. Thus, we prove the following Lemma:
\begin{lemma}
    \label{lemma:continuity-of-predictor}
    Let Assumption \ref{assumption:lipschitz-dynamics} hold. Then, for any $X_1, X_2 \in \mathcal{X}$, $U_1, U_2 \in C^2([0, 1]; \mathcal{U})$, and $\varphi_1, \varphi_2 \in (\underline{D}, \overline{D})$ the predictor operator $\breve{\mathcal{P}}$ given in Definition \ref{defintion:predictor-operator} satisfies 
    \begin{align} \label{eq:predictor-lipschitz-bound}
        \|\breve{\mathcal{P}}(&X_1, U_1, \varphi_1) - \breve{\mathcal{P}}(X_2, U_2, \varphi_2) \|_{L^\infty[0, 1]} \nonumber \\   &\leq C_{\breve{\mathcal{P}}} \left(|X_1-X_2| + \|U_1 - U_2\|_{L^\infty[0, 1]} + |\varphi_1 - \varphi_2| \right)\,,
    \end{align}
    with Lipschitz constant
    \begin{align}
        C_{\breve{\mathcal{P}}} &= e^{\overline{D}C_f} \max\left\{1, \Xi, \overline{D}C_f\right\}\,, \\
        \Xi &= C_f\left[\overline{U} + e^{\overline{D}C_f}(\overline{X} + C_f\overline{D}\overline{U})\right]\,. 
    \end{align}
\end{lemma}
\begin{pf}
    For all $s \in [0, 1]$, let $\bar{P}_1(s) := \breve{\mathcal{P}}(X_1, U_1, \varphi_1)$ and likewise  $\bar{P}_2(s) := \breve{\mathcal{P}}(X_2, U_2, \varphi_2)$. Note that, for $s\in [0,1]$ the predictor is uniformly bounded. First, by definition and Assumption \ref{assumption:lipschitz-dynamics}, we have 
    \begin{align}
        \bar{P}_1(s) &= X_1 + \varphi_1 \int_0^s f(\bar{P}_1(\theta), U_1(\theta)) d \theta \nonumber \\ 
        &\leftstackrel{\eqref{eq:lipschitz-dynamics-cond}}{\leq} X_1 + \varphi_1 \int_0^s C_f(|\bar{P}_1(\theta)| + |U_1(\theta)|) d\theta \nonumber  \\ 
        &\leq X_1 + \varphi_1 C_f \|U_1\|_{L^\infty[0, 1]} + \varphi_1 C_f \int_0^1 |\bar{P}_1(\theta)| d \theta 
    \end{align}
    Applying Gronwall's inequality yields 
    \begin{align} \label{eq:predictor-uniform-bound}
        |\bar{P}_1(s)| \leq e^{\overline{D}C_f}(\overline{X} + C_f \overline{D} \phantom{,} \overline{U})\,. 
    \end{align}
    Then, applying \eqref{eq:predictor-uniform-bound} yields the following calculation
       \begin{alignat}{2}
        \bar{P}_1(s) - \bar{P}_2(s) &=&& X_1 - X_2 + \varphi_1 \int_{-1}^s f(\bar{P}_1(\theta ), U_1(\theta )) d \theta  \nonumber  \\ & &&\nonumber  - \varphi_2 \int_{-1}^s f(\bar{P}_2(\theta ), U_2(\theta )) d\theta \\ \nonumber
        &\leq&& |X_1-X_2| \nonumber \\\nonumber & && + (\varphi_1-\varphi_2) \int_{-1}^s  f(\bar{P}_1(\theta ), U_1(\theta )) d\theta \\\nonumber & && + \varphi_2 \int_{-1}^s\bigg( f(\bar{P}_1(\theta), U_1(\theta )) \\ \nonumber  & &&- f(\bar{P}_2(\theta), U_2(\theta )) \bigg) d \theta   \\ 
        &\leftstackrel{\eqref{eq:lipschitz-dynamics-cond}}{\leq}&& |X_1 - X_2| + (\varphi_1-\varphi_2) C_f \|U_1\|_{L^\infty[0, s]} \nonumber \\ \nonumber & &&  + (\varphi_1-\varphi_2) C_f \int_{-1}^s |P(\theta )| d \theta \\ & && + \varphi_2C_f \int_{-1}^s |\bar{P}_1(\theta) - P_2(\theta)|\nonumber \\ & && \nonumber + |U_1(\theta)-U_2(\theta)| d \theta \\ 
        &\leftstackrel{\eqref{eq:predictor-uniform-bound}}{\leq}&&  |X_1 - X_2| \nonumber \\ & &&  + |\varphi_1-\varphi_2| C_f \bigg[\overline{U}  + e^{C_f\overline{D}}(\overline{X} + C_f \overline{D} \phantom{,} \overline{U})\bigg] \nonumber \\ & && + \overline{D}C_f \|U_1-U_2\|_{L^\infty[-1, s]} \nonumber \\ & && 
        + \overline{D} C_f \int_{-1}^s |\bar{P}_1(\theta) - \bar{P}_2(\theta)|d\theta \,.
    \end{alignat}
    Applying Gronwall's again yields the desired result. 
\end{pf}

Given the continuity in Lemma \ref{lemma:continuity-of-predictor}, we can then apply Theorem \ref{thm:neural-operator-uat} to the predictor operator yielding the existence of an arbitrarily close neural operator approximation:
% \begin{theorem} \label{thm:uat-predictor}
%     Fix a compact set $K \subset \mathcal{X} \times \mathcal{X}' \times C^2([0, 1]; \mathcal{U}) \times C^1([0, 1]; \mathcal{U}') \times \mathcal{D} \times \mathcal{D'}$. Then, for all $\overline{X}$, $\overline{X}'$, $\overline{U}$, $\overline{U}'$, $\overline{D}$, $\overline{D}'$, $\epsilon > 0$, there exists a neural operator approximation $\hat{\mathscr{M}} : K \to C^1([0, 1]; \mathbb{R}^n) \times C^0([0, 1]; \mathbb{R}^n)$ such that 
%     \begin{align}
%         \sup_{x_1, x_2 \in K} &|\mathscr{M}(x_1, x_2)(s)  - \hat{\mathscr{M}}(x_1, x_2)(s) |  < \epsilon \,, 
%     \end{align}
%     or, equivalently
%     \begin{align}
%         \sup_{a \in K} &|\breve{P}(a)(s) - \hat{P}(a)(s) | \nonumber \\ &+
%          \sup_{(a, b) \in K} |\breve{P}'
%          (a, b)(s) - \hat{P}'(a, b)(s) |< \epsilon \,, 
%     \end{align}
%     for all $s \in [0, 1]$ where (with slight abuse of shorthand notation) $a = (X, U, \varphi)$ and $b = (X', U', \varphi')$.  
% \end{theorem}
\begin{theorem} \label{thm:uat-predictor}
    Fix a compact set $K \subset \mathcal{X} \times C^2([0, 1]; \mathcal{U}) \times \mathcal{D}$. Then, for all $\overline{X}$, $\overline{U}$, $\overline{D}$, $\epsilon > 0$, there exists a neural operator approximation $\breve{\mathcal{P}}_{\text{NO}} : K \to C^1([0, 1]; \mathbb{R}^n)$ such that 
    \begin{align}
        \sup_{(X, U, \varphi) \in K} &|\breve{\mathcal{P}}(X, U, \varphi)(s)  - \breve{\mathcal{P}}_{\text{NO}}(X, U, \varphi)(s) |  < \epsilon \,, 
    \end{align}
    for all $s \in [0, 1]$.
\end{theorem}

Now, we have established the existence of an arbitrarily close approximate predictor, but we did not discuss the detail on the number of parameters or data required to obtain this bound. Such analysis is beyond the goal of this paper, but we refer the reader to \cite{mukherjee2024size}, \cite{NLM2024data}, \cite{LST2024discretization} for further details. Further, given the approximation is arbitrarily close in $\epsilon$, this perturbation will affect the stability of the system compared to the global asymptotic result conducted with the exact predictor (which can never be implemented in practice). Therefore, in the next section, we identify the exact effect of the approximate predictor on the overall feedback loop. 
\section{Stability analysis under approximate predictors} \label{sec:stability}
We are now ready to analyze the plant \eqref{eq:plant-ode-pde-1}, \eqref{eq:plant-ode-pde-2}, \eqref{eq:plant-ode-pde-3} under the controller with the neural operator approximated predictor 
\begin{align} \label{eq:approximate-predictor-feedback-law}
    U(t) =& \kappa (\breve{P}_{\text{NO}}(t))\,, \\
    \breve{P}_{\text{NO}}(t) \coloneq & \breve{\mathcal{P}}_{\text{NO}}(X(t), T_{D}(t)U, \breve{D}(t))(1)\,, 
\end{align}
where $T_{D}(t)$ is the historical control operator which applied to a function $U$ yields
\begin{align} \label{eq:shift-operator-control-history}
    (T_D(t)U) (x) := U(t-D(x -1))\,,  \quad  x \in [0, 1)\,.
\end{align} 
The operator $T_{D}(t)$ shifts the function $U$ back by $t-D(x-1)$ units such that, we have $u(x, t) = (T_{D}(t)U)(x)$ and therefore the resulting output of the exact predictor operator satisfies \eqref{eq:exact-predictor-adaptive}. Notice that, in this case, we consider the full actuator measurement such that $u(x, t)$ is always known which is not necessarily guaranteed for every application. However, as shown in Section \ref{sec:numerical}, this is a reasonable assumption for biological systems of which we study in this work and in \ref{sec:unmeasured}, we also present analysis for the unmeasured case. 

We are now ready to present the main result of the paper. 
\begin{theorem} \label{thm:main-result}
    Let the system \eqref{eq:plant-ode-pde-1}, \eqref{eq:plant-ode-pde-2}, \eqref{eq:plant-ode-pde-3} satisfy Assumptions \ref{assumption:strongly-forward-complete}, \ref{assumption:gas}, \ref{assumption:lipschitz-dynamics}, \ref{assumption:growth-condition}. Define the functional
    \begin{align}
        \Gamma(t) = |X(t) |^2 + \int_{t-D}^t U(\theta)^2d \theta + |\tilde{D}(t)|^2\,,  
    \end{align}
    where $\tilde{D}(t) = D-\breve{D}(t)$. Then, there exists constants $\gamma^\ast$, $b^\ast(\bar{D}, C_f)$, $\overline{\Gamma}(\overline{X}, \overline{U}, \overline{D}) > 0$, $\beta_1^\ast \in \mathcal{KL}$ and class $\mathcal{K}_\infty$ functions $\alpha_{1:5}^\ast$ such that if $\gamma < \gamma^\ast$, $b > b^\ast$, $\epsilon < \epsilon^\ast$ where
    \begin{align} \label{eq:epsilon-star}
        \epsilon^\ast = (\alpha_1^\ast)^{-1}(\overline{\Gamma
     }-\alpha_2^\ast(\overline{\Delta D}))\,, 
    \end{align}
    where $\overline{\Delta D} = \overline{D}- \underline{D}$ and the initial state is constrained to
    \begin{align} \label{eq:gamma-zero}
    \Gamma(0) \leq \alpha_3^\ast(\overline{\Gamma} - \alpha_1^\ast(\epsilon) - \alpha_2^\ast(\overline{\Delta D}))\,, 
    \end{align}
    then, 
    \begin{align} \label{eq:gamma-t}
        \Gamma(t) \leq \alpha_3^\ast(\Gamma(0)) + \alpha_1^\ast(\epsilon) + \alpha_2^\ast(\overline{\Delta D})\,, 
    \end{align}
    and 
    \begin{align} \label{eq:regulation}
       |X(t)|^2 &\leq \beta_1^\ast(\Gamma(0), t) + \alpha_4^\ast(\overline{\Delta D}) + \alpha_5^\ast (\epsilon)\,,  \\ 
        \|u(t)\|_{L^2[0, D]}^2 &\leq \beta_1^\ast(\Gamma(0), t) + \alpha_4^\ast(\overline{\Delta D}) + \alpha_5^\ast(\epsilon)\,.   
    \end{align}
\end{theorem}

Notice that Theorem \ref{thm:main-result} is weaker than the result in \cite[Theorem 7]{bhan2023neural} because it depends on the delay projection bounds $\overline{D}$, $\underline{D}$. This is expected since the error introduced by the predictor is additive, leading to complications similar to those in robust adaptive control, where regulation depends on projector bounds (See \cite{IKHOUANE1998429}). Additionally, it may seem counterintuitive that $\epsilon^\ast$ increases with $\overline{\Gamma}$, but this is expected: a larger radius of states allows for a larger transient, thus loosening the $\epsilon^\ast$ bound. 

Finally, we compare Theorem 3 with other adaptive neural-operator PDE control results. In \cite[Theorem 4]{bhan2024adaptivecontrolreactiondiffusionpdes} and \cite[Theorem 4]{lamarque2024adaptiveneuraloperatorbacksteppingcontrol}, the gain kernel is approximated, introducing a multiplicative error and yielding global stability, but requiring approximation of the update parameter’s time derivative. In contrast, Theorem \ref{thm:main-result} approximates $\breve{P}$—analogous to the direct control law as in \cite[Theorem 3]{bhan2023neural}—producing an additive error and a semi-global practical result without the derivative. This highlights the trade-off between global and semi-global guarantees based on which controller component is approximated.
\begin{pf}
    We first introduce a backstepping transform, that under the feedback law \eqref{eq:approximate-predictor-feedback-law}, transforms the system into a target system that we then use for our stability analysis. 
    \begin{lemma} \label{lemma:backstepping-lemma}
    Under the backstepping transformation with the \emph{exact predictor} given by
    \begin{align} \label{eq:lemma2-bcks-transform}
        w(x, t) = u(x, t)- \kappa(\breve{p}(x, t))\,,
    \end{align}
    the system \eqref{eq:plant-ode-pde-1}, \eqref{eq:plant-ode-pde-2}, \eqref{eq:plant-ode-pde-3} with control law given by \eqref{eq:approximate-predictor-feedback-law}
    \begin{subequations} is transformed into    
    \begin{align}
        \dot{X}(t) &= f(X(t), \kappa(X(t)) + w(0, t))\,, \label{eq:plant-target-ode-pde-1} \\ 
        Dw_t &= w_x - \tilde{D}(t) q_1(x, t) - D \dot{\breve{D}}(t) q_2(x, t)\,,\label{eq:plant-target-ode-pde-2}  \\
        w(1, t) &= \kappa(\breve{p}(x, t)) - \kappa(\breve{p}_{\text{NO}}(x, t))\,, \label{eq:plant-target-ode-pde-3} 
    \end{align}
    \end{subequations}
    where $\tilde{D} = D-\breve{D}(t)$, \newline  $\breve{p}_{\text{NO}}(x, t) = \breve{\mathcal{P}}_{NO}(X(t), T_D(t)U, \hat{D}(t))(x)$, and $q_2$ is given by
    \begin{align} \label{eq:q2}
        q_2 =& \frac{\partial \kappa}{\partial \breve{p}}(\breve{p}(x, t)) \int_0^x \Phi(x, y, t) f(\breve{p}(y, t), \kappa(\breve{p}(y, t)) \nonumber \\ &+w(y, t)) dy\,. 
    \end{align}
    \end{lemma}
    Since we used the exact backstepping transform in \eqref{eq:lemma2-bcks-transform}, the proof of Lemma \ref{lemma:backstepping-lemma} is exactly the same as the proof of \cite[Lemma 1]{delphine} except the boundary term at $x=1$ which is obtained via direct substitution of $U(t) =\kappa(\breve{P}_{NO}(\cdot)) - \kappa(\breve{P}(\cdot))$ in \eqref{eq:lemma2-bcks-transform}.  

    To analyze \eqref{eq:plant-target-ode-pde-1}, \eqref{eq:plant-target-ode-pde-2}, \eqref{eq:plant-target-ode-pde-3}, we introduce the following Lyapunov-Krasovskii functional
    \begin{align}
        W(t) &= D\log N(t) + \frac{b}{\gamma}\tilde{D}(t)^2  \\ 
        N(t) &= 1+V(x) + b\int_0^1(1+x)w(x, t)^2 dx\,.
    \end{align}
    Taking the time derivative and substituting $w_t$ yields
    \begin{align}
        \dot{W}(t) =& \frac{1}{N(t)} \bigg( D \frac{\partial V(X)}{\partial X} \bigg[f(X(t), \kappa(X(t)) + w(0, t)) \nonumber  \bigg] \bigg) \nonumber \\ & + \frac{2b}{N(t)} \int_0^1 (1+x) w(x, t)\bigg(w_x(x, t) - \tilde{D}(t) q_1(x, t) \nonumber \\ &- D \dot{\breve{D}}(t) q_2(x, t) \bigg) dx - \frac{2b}{\gamma} \tilde{D} \dot{\breve{D}}(t) \,.
    \end{align}
    Using the Lipschitzness of $f$ in Assumption \ref{assumption:lipschitz-dynamics} and the Lyapunov condition in Assumption \ref{assumption:gas}, we have
    \begin{align}
        \dot{W}(t) &\leq  \frac{1}{N(t)} \left(-D \lambda |X(t)|^2 + DC_2 C_f |X(t)| |w(0, t)| \right) \nonumber \\  & + \frac{2b}{N(t)} \int_0^1 (1+x) w(x, t)\bigg(w_x(x, t) - \tilde{D}(t) q_1(x, t) \nonumber \\ &- D \dot{\breve{D}}(t) q_2(x, t) \bigg) dx - \frac{2b}{\gamma} \tilde{D} \dot{\breve{D}}\,.
    \end{align}
    Distributing the second term, applying integration by parts, and substituting \eqref{eq:update-law} yields
    \begin{align}
        \dot{W}(t)  \leq& \nonumber  \frac{1}{N(t)} \bigg( -D \lambda |X(t)|^2 + DC_2C_f |X(t)||w(0, t)| \nonumber \\  &+ bw^2(1, t) - bw^2(0, t)  - b\|w(t)\|_2^2 \nonumber \\ &+ 2bD \dot{\breve{D}}(t) \int_0^1 (1+x)w(x, t) q_2(x, t) dx  \bigg) \nonumber \\ &+ \frac{2b}{\gamma}\tilde{D}(t)(\gamma \phi(t) - \dot{\breve{D}}(t))\,.
    \end{align}
    Using \cite[Lemma 2 and 3]{delphine} which states that there exists constants $M_6, M_6, M_7 > 0$ such that 
    \begin{align}
     M_6(|X| + \|w(t)\|_2)  \geq& |\breve{p}(x, t)|\,, \label{eq:breve-preidctor-bound} \\ 
     M_6\left(|X|^2 + \|w(t)\|_2^2\right) \geq&  2bD \bigg|\int_0^1 (1+x) \nonumber \\ & \times q_2(x, t) w(x, t) dx \bigg|\,, \\
    \gamma M_7 \geq &\left| \dot{\breve{D}}(t) \right|\,, 
\end{align}
in conjunction with Young's inequality yields
\begin{align}
    \dot{W}(t) \leq & \nonumber \frac{1}{N(t)} \bigg(-\eta \left(|X(t)|^2 + \|w(t)\|_2^2 \right) \\ \nonumber  &- \bigg(b- \frac{C_2^2 D C_f^2}{2 \lambda}\bigg) w^2(0, t) \\ &+ \gamma M_6 M_7 \bigg(|X(t)|^2 + \|w(t)\|_2^2 \bigg)  \nonumber  \\ & + b w^2(1, t)\bigg)\,,
\end{align}
where $\eta := \min(D\lambda/2, b)$. Thus, given that 
\begin{align}
    b &> \frac{C_2^2 \overline{D} C_f^2 }{2\lambda} =: b^\ast\,,  \\
    \gamma &<  \frac{\eta}{M_6M_7} =: \gamma^*\,, 
\end{align}
guarantees there exists $C>0$ such that
\begin{align} \label{eq:w-dot}
    \dot{W}(t) \leq& \nonumber  -\frac{C}{N(t)}(|X(t)|^2 + \|w(t)\|^2_2) \\ & + \frac{b}{N(t)}(\kappa(\breve{p}(x, t)) - \kappa(\breve{p}_{\text{NO}}(x, t)))^2 \,.
\end{align}
Now, using the definition of $W(t)$ and Assumption \ref{assumption:gas}, we have that 
\begin{align}
    e^{W(t)/D - \frac{b}{\gamma{D}}(\tilde{D}(t))^2} - 1 \leq& \max\{C_1, 2b\} \nonumber \\ & \times (|X(t)|^2 + \|w(t)\|_2^2)\,,
\end{align}
Using reverse Young's inequality (See Appendix \ref{appendix:reverse-youngs}), we have
\begin{align}
    e^{W(t)/D}e^{-\frac{b}{\gamma{D}}(\tilde{D}(t))^2} \geq 2e^{W(t)/(2D)}-e^{\frac{b}
    {\gamma D}{\tilde{D}(t))^2}} 
\end{align}
yielding
\begin{align}
 2e^{W(t)/D} - &e^{\frac{b}{\gamma D}(\tilde{D}(t))^2}-1 \nonumber \\ 
 &\leq \max\{C_1, 2b\}(|X(t)|^2 + \|w(t)\|_2^2)\,, 
\end{align}
which then results in 
\begin{align}
   - (|X(t)|^2 + \|w(t)\|_2^2) \leq&  \frac{1}{\max\{C_1, 2b\}} \bigg(-2e^{W(t)/D} \nonumber \\ &+ e^{\frac{b}{\gamma D}(\tilde{D}(t))^2} + 1 \bigg)\label{eq:w-lowerbound}\,. 
\end{align}
Substituting into \eqref{eq:w-dot} yields
\begin{align}
    \dot{W}(t) \leq& -\frac{C}{\max\{C_1, 2b\}N(t)}   \nonumber 2e^{\frac{W(t)}{D}} \\ &+ \bigg[ \frac{C}{\max\{C_1, 2b\}N(t)}\left(1+e^{\frac{b}{D\gamma}(\overline{\Delta D})^2}\right) \nonumber \\  &+ b(\kappa(P(x, t)) - \kappa(\hat{p}(x, t)))^2 \bigg]\,. 
\end{align}
Then, by Assumption \ref{assumption:growth-condition}, we have that for all $x \in [0, 1]$, $t\geq 0$
\begin{align}
    (\kappa(P(x, t)) - \kappa(\hat{p}(x, t)))^2 \leq& M_3^2 |P(x, t) - \hat{p}(x, t)|^2 \nonumber \\ \leq& M_3^2 \epsilon^2\,.
\end{align}
Following %the argument of 
\cite[Theorem C.3]{kkk}, we have that there exists functions $\beta_1 \in \mathcal{KL}$ and $\alpha_{1:3}\in \mathcal{K}_\infty$ such that 
\begin{align}
    W(t) \leq& \beta_1(W(0), t) +\nonumber  \alpha_1(\overline{\Delta D})  \\ &+  \alpha_2(\sup_{0 \leq \tau \leq t}(\kappa(\breve{p}(x, t)) - \kappa(\breve{p}_{\text{NO}}(x, t)))^2 ) \nonumber \\ 
    \leq& \beta_1(W(0), t) + \alpha_1(\overline{\Delta D}) + \alpha_3(\epsilon)\label{eq:w-stability-estimate}\,,
\end{align}
when $\epsilon$ and $\overline{\Delta D}$ is small enough relative to all possible values of $W$ (i.e. $\epsilon + (\overline{\Delta D}) \leq \alpha_4(W(t)) \leq \alpha_5(\overline{X}+\overline{U}+\overline{D})$ where $\alpha_4, \alpha_5 \in \mathcal{K}_\infty$). 
 Now, \eqref{eq:breve-preidctor-bound} and Assumption \ref{assumption:growth-condition} implies that there exists constants $r_1, r_2, s_1, s_2 > 0$ such that
\begin{align}
    \|u(t)\|_2^2 \leq& r_1|X(t)|^2 + r_2\|w(t)\|^2\,,  \label{eq:u-xw-bound} \\
    \|w(t)\|_2^2 \leq& s_1|X(t)|^2 + s_2\|u(t)\|^2\,.\label{eq:w-ux-bound}
\end{align}
To obtain an estimate on $\Gamma$, note that from \cite[Eqn. (47), (48)]{delphine}, we have that 
\begin{align}
    \Gamma(t) &\leq \left(D\left(r_1 + \frac{r_2}{b} \right) + 1 + \frac{\gamma D}{b} \right) \left(e^{\frac{W(t)}{D}} - 1 \right) \,,\label{eq:gamma-upperbound} \\
    W(t) &\leq D \left(C_1 + 2b \left(s_1 + \frac{s_2}{D} \right) + \frac{b}{\gamma D} \right) \Gamma(t)\,.
\end{align}
Therefore, substituting in the stability estimate in \eqref{eq:w-stability-estimate}, we obtain the exists of class $K^\infty$ functions $\alpha_1^\ast$, $\alpha_2^\ast$, $\alpha_3^\ast$ such that 
\begin{align}
    \Gamma(t) \leq \alpha^\ast_1(\Gamma(0)) + \alpha_2^\ast(\epsilon)+\alpha_3^\ast(\overline{\Delta D}) \,,
\end{align}
when $\Gamma(0) \leq (\alpha^\ast_1)^{-1}(\overline{\Gamma} - \alpha_2^\ast(\epsilon) - \alpha_3^\ast(\overline{\Delta D}))$ where $\overline{\Gamma} = \overline{X}+ \overline{U} + \overline{D}$. 

To show obtain a bound on $X(t)$, note that we have
\begin{align}
    |X(t)|^2 \leq V(t) \leq (e^{\frac{W(t)}{D}}-1)\, \label{eq:x-upperbound}.
\end{align}
Using the stability estimate on $W(t)$ in \eqref{eq:w-stability-estimate}, along with the inequalities \eqref{eq:x-upperbound}, we have that there exists $\beta_2 \in \mathcal{KL}$, $\alpha_6, \alpha_7 \in \mathcal{K}_\infty$ such that
\todo{fix}
\begin{align}
    |X(t)|^2 \leq \beta_2(W(0), t) + \alpha_6(\overline{\Delta D}) + \alpha_7 (\epsilon)\,. 
\end{align}

Similarly, to show a stability estimate on $U(t)$, we use \eqref{eq:u-xw-bound}, \eqref{eq:w-ux-bound}, along with the definition of $W(t)$, we have that 
\begin{align}
    \int_{t-D}^t U(s)^2ds = D\|u(t)\|_2^2 &\leq D(r_1 |X(t)|^2 + r_2\|w(t)\|_2^2)\nonumber  \\
    &\leq D\left(r_1 + \frac{r_2}{b}\right) \left(e^{\frac{W(t)}{D}} - 1\right)\,. \label{eq:u-upper-bound}
\end{align}
Then, repeating the argument above with the stability estimate in \eqref{eq:w-stability-estimate}, we have there exists $\beta_3 \in\mathcal{KL}$, $\alpha_8, \alpha_9 \in \mathcal{K}_\infty$ such that 
\begin{align}
    \|u(t)\|_{L^2[0, D]}^2 \leq \beta_3(W(0), t) + \alpha_8(\overline{\Delta D}) + \alpha_9(\epsilon)\,.   
\end{align}
Letting $\beta_1^\ast(\cdot, \cdot) = \max\{\beta_2(\cdot, \cdot), \beta_3(\cdot, \cdot)\}$, $\alpha_4^\ast(\cdot) = \max\{\alpha_6(\cdot), \alpha_8(\cdot)\}$ and $\alpha_5^\ast(\cdot) = \max\{\alpha_7(\cdot), \alpha_9(\cdot)\}$ and applying the inequality \eqref{eq:gamma-upperbound} completes the result.

\end{pf}
\section{Numerical Experiments} \label{sec:numerical}

\begin{figure*}[!htbp]
    \centering
    \includegraphics{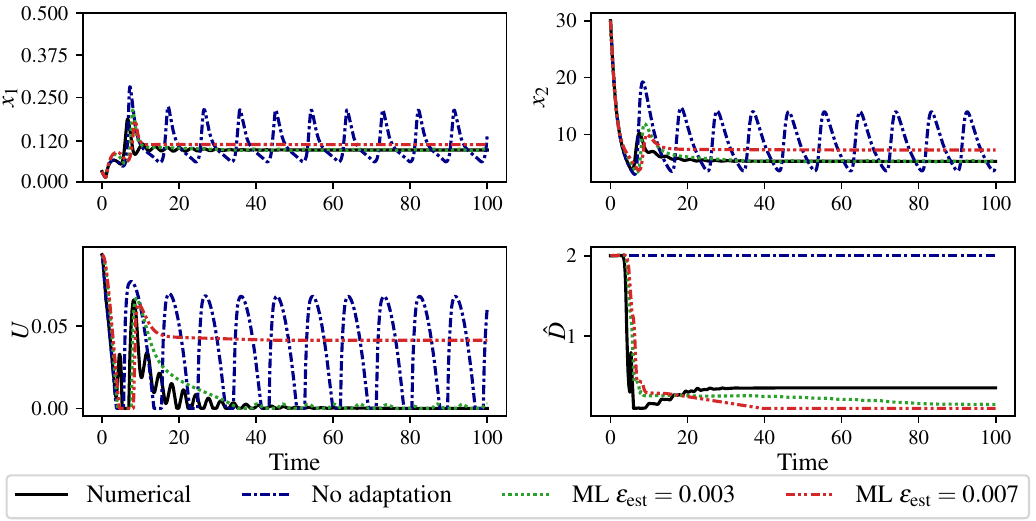}
    \caption{Simulation of the plant \eqref{eq:main-problem} with various approximate predictors. The initial state is $X(0) = [0.03, 30]$, delay is $D=1$, $\hat{D}(0) = 2$, $\gamma=1000$, $b=1$. The black line indicates the numerical predictor, the blue-line is the predictor without delay-adaptation, the red line indicates the DeepONet predictor with higher error and the green line indicates the DeepONet predictor trained to optimality.}
    \label{fig:main-fig}
\end{figure*}

For the simulation analysis, we provide all code, numerical parameters, and datasets publicly on Github (\url{https://github.com/lukebhan/Neural-Operator-Delay-Adaptive-Predictor-Feedback}).
\color{black}

For numerical validation, we consider the biological system in \cite{4282275} consisting of two proteins. The activator protein that promotes expression of itself and a repressor protein that represses the expression of the activator. Such biological clocks play fundamental roles in cell physiology \cite{Pomerening2003} and control of such systems is of valuable interest to synthetic biologists for the design of new medicines. The system dynamics manifest as 
\begin{align}
    \dot{x}_1 &= -x_1 + f_1(x_1, x_2) + U(t-D)\,, \\
    \dot{x}_2 &= -\frac{x_2}{2} + f_2(x_1)\,,
\end{align}
where $x_1$ is the concentration of the activator protein, $x_2$ is the concentration of the repressor protein and $f_1$, $f_2$ are Hill functions given by 
\begin{align}
    f_1(x_1, x_2) &= \frac{K_1x_1^2 + K_a}{1+x_1^2 + x_2^2}\,, \\
    f_2(x_1) &= \frac{K_2x_1^2 + K_b}{1+x_1^2}\,,
\end{align}
with $K_1=K_2 = 300, K_a=0.04, K_b=0.004$. 
The input $U$ controls the activator protein concentration with a delay due to the time to pass through an inlet pipe. Additionally, we assume the pipe is equipped with a UV spectrometer such that we can measure the presence of the activator protein along the pipe and thus the distributed input $u(\cdot, t)$ is known. 

A control law satisfying Assumption \ref{assumption:gas} is given by
\begin{align}
    \kappa(X(t)) = -f_1(x_1, x_2) + f_1(x^\ast_1, x_2^\ast)\,,
\end{align}
where $(x_1^\ast, x_2^\ast) = (0.0939, 5.2525)$ is an unstable equilibrium setpoint of the system. The associated Lypaunov function is given by $V = (X-X^\ast)^T(X-X^\ast)$ and note the system experiences a limit cycle \cite[Figure 2]{delphine} when an open-loop control law $U(\cdot) = 0$ is used.  

To train a neural operator approximate predictor, we generate a dataset of $5000$ instances by simulating the system with a numerical predictor \cite{iassonBook} across various initial conditions and delay estimates (see code for parameters; dataset generation took 60 minutes). Two neural operator architectures (DeepONet, FNO) are trained in approximately 10 minutes on an NVIDIA 3090Ti GPU. Both perform similarly, though DeepONet achieves a larger speedup (Table \ref{tab:comp-time}).

Figure \ref{fig:main-fig} shows system simulations using the DeepONet predictor. To illustrate the effect of $\epsilon$, we compare two predictors: one trained with early stopping (large $\epsilon$) and one trained to optimality. As shown by the red line, early stopping results in convergence around the equilibrium with a larger radius. Even the optimally trained predictor, though not visually apparent, does not fully reach the setpoint, converging to $(0.094,5.39)$ instead of $(0.0939,5.2525)$ due to the uniform approximation error $\epsilon$. In both cases, $\epsilon$ directly determines the convergence radius, consistent with Theorem \ref{thm:main-result}.

Finally, we evaluated the computational speedup of the neural operator predictor. Across all discretizations, the ML predictor outperforms the numerical approach, achieving up to $15\times$ speedup at $dx=0.001$, with stability requiring $dx<0.005$ (Euler scheme). We note this speedup improvement is lower than the $100\times$ reported in \cite[Table 1]{pmlr-v283-bhan25a} due to the simpler dynamics in this example; however, applying the same approach to more computationally expensive forward dynamics would likely yield even greater gains with DeepONet.
\begin{table}[htbp]
\centering
\caption{Computation time for various approximate predictors averaged over $1000$ samples for the biological proetin example (seconds). }
\label{tab:comp-time}
\resizebox{\linewidth}{!}{%
\begin{tabular}{l|ccccc}
\begin{tabular}[c]{@{}l@{}}Step \\ size (dx)\end{tabular} & \multicolumn{1}{l}{Numerical $\downarrow$} & \multicolumn{1}{l}{DeepONet $\downarrow$} & \multicolumn{1}{l}{FNO $\downarrow$} & \multicolumn{1}{l}{\begin{tabular}[c]{@{}l@{}}DeepONet\\  Speedup $\uparrow$\end{tabular}} & \multicolumn{1}{l}{\begin{tabular}[c]{@{}l@{}}FNO\\ Speedup $\uparrow$\end{tabular}} \\ \hline
0.01                                                      & 1.601                                      & \textbf{0.496}                            & 1.331                                & \textbf{3.22x}                                                                             & 1.20x                                                                                \\
0.005                                                     & 3.295                                      & \textbf{0.587}                            & 1.440                                & \textbf{5.61x}                                                                             & 2.29x                                                                                \\
0.001                                                     & 18.197                                     & \textbf{1.212}                            & 2.108                                & \textbf{15.01x}                                                                            & 8.63x                                                                               
\end{tabular}%
}
\end{table}

\section{Control in presence of an unmeasured distributed input} \label{sec:unmeasured}
In Section \ref{sec:stability}, we assumed the actuator state $u(x,t)$ was known. In many real-world systems—biological feedback loops, vaccine distribution, or economic interventions—the actuator is often unknown and subject to uncertain or unmeasured delays. Here, we extend Section \ref{sec:stability} to unknown actuator states, starting with the definition of the actuator state estimate:
\begin{align}
    \hat{u}(x, t) = U(t+\hat{D}(t)(x-1))\,, \quad x\in [0, 1]\,, 
\end{align}
where $\hat{D}$ is an estimate of the delay. 
The control law is now computed as 
\begin{align}
    U(t) &= \kappa(\hat{p}(1, t))\,, \\
    \hat{p}(x, t) &= X(t+\hat{D}(t)x)\nonumber  \\ &= X(t) + \hat{D}(t) \int_0^x f(\hat{p}(y, t), \hat{u}(y, t)) dy \,,
\end{align}
where the predictor $\hat{p}$ is computed using the approximate actuator state $\hat{u}(x,t)$. Following \cite{delphine}, we assume the delay update law satisfies:
\begin{assumption} \label{assumption:delay-func}
    There exists a $C^1$ function $\hat{\phi}$, a positive parameter $\gamma > 0$ and class $\mathcal{K}$ functions $\rho_1$, $\rho_2$ such that 
    \begin{align}
        \dot{\hat{D}}(t) = \gamma \text{Proj}_{[\underline{D}, \overline{D}]}\left\{\hat{D}(t)\,, \hat{\phi}(t)\right\}\,,  
    \end{align}
    where $\hat{\phi}(t)$ satisfies
    \begin{align}
        |\hat{\phi}(t)| \leq \rho_1(\Xi(t)) \text{ and } |\dot{\hat{\phi}}(t)|\leq \rho_2 (\Xi(t))
    \end{align}
    with the function $\Xi(t)$ given by 
    \begin{align}
        \Xi(t) =|X(t)| &+ \int_{t-\max\{D, \hat{D}(t)\}}^t |U(s) |ds \nonumber \\ 
        &+\int_{t-\max\{D, \hat{D}(t)\}}^t |\dot{U}(s)| ds \nonumber \\ &+\int_{t-\max\{D, \hat{D}(t)\}}^t \left|\ddot{U}(s)\right|ds\,. 
    \end{align}
\end{assumption}
For example, we will consider the update law given by
\begin{align}
    \hat{\phi}(t) \nonumber =& 2\sgn (\hat{w}_x(1, t)) q_3(1, t) \\ &+ \int_0^1 (1+x)\bigg[ q_3(x, t) \sgn (\hat{w}(x, t)) \nonumber \\ & + q_4(x, t)\sgn (\hat{w}_x(x, t)) \bigg]dx \,,
\end{align}
where the backstepping transform $\hat{w}(x, t)$ is as expected with the estimated actuator state
\begin{align}
    \hat{w}(x, t) = \hat{u}(x, t) - \kappa(\hat{p}(x, t))\,,
\end{align}
and the functions $q_3$, $q_4$ are given by
\begin{align}
    q_3(x, t) &= \frac{d \kappa}{d\hat{p}}(\hat{p}(x, t)) f(\hat{p}(0, t), \hat{u}(0, t))\,, \\
    q_4(x, t) &= \frac{\partial }{\partial x} q_3(x, t)\,. 
\end{align}

We are now ready to state the our main result. 
\begin{theorem} \label{thm:main-result-no-measurement}
    Consider the plant \eqref{eq:plant-ode-pde-1}, \eqref{eq:plant-ode-pde-2}, \eqref{eq:plant-ode-pde-3} satisfying Assumptions \ref{assumption:strongly-forward-complete}, \ref{assumption:gas},  \ref{assumption:lipschitz-dynamics}, \ref{assumption:growth-condition}, \ref{assumption:delay-func} and the functional 
    \begin{align}
        \Upsilon(t) = |X(t)| &+ \int_{t-\max\{D, \hat{D}(t)\}}^t |U(s)|ds \nonumber \\ & + \int_{t-\max\{D, \hat{D}(t)\}}^t |\dot{U}(s)| ds \nonumber \\ & +\int_{t-\max\{D, \hat{D}(t)\}}^t |\ddot{U}(s)| ds + \tilde{D}(t)^2  \,.
    \end{align}
    Then, there exists functions $\beta_2^* \in \mathcal{KL}$ and $\alpha_6^\ast, \alpha_7^\ast, \alpha_8^\ast, \alpha_9^\ast  \in \mathcal{K}_\infty $ and constants $\overline{\Upsilon} > 0$, $\gamma^\ast(\Upsilon(0))>0$, 
    \begin{align}
        \hat{\epsilon}^*(\overline{\Upsilon}) \coloneq (\alpha_6^\ast)^{-1}(\overline{\Upsilon})> 0\,, 
    \end{align} and a region of attraction 
    \begin{align}
        R_\Upsilon(\overline{\Upsilon}, \epsilon) \coloneq \alpha_7^{\ast}(\overline{\Upsilon} - \alpha_6^\ast(\epsilon)) > 0\,, 
    \end{align} such that if $\gamma < \gamma^*$, $\epsilon <\hat{\epsilon}^\ast$, and the initial state is constrained to $|\Upsilon(0) |\leq R_\Upsilon$,  
    then, the feedback law $U(t) = \kappa(\hat{p}_{\text{NO}}(1, t))$ with the estimator 
    \begin{align}
        \dot{\hat{D}}(t) = \gamma\text{Proj}_{\underline{D}, \overline{D}]}\left\{\hat{D}(t), \hat{\phi}(t) \right\}\,, 
    \end{align}
    guarantees for all solutions
    \begin{align}
        \Upsilon(t) \leq \alpha_7^\ast  (\Upsilon(0)) + \alpha_6^\ast(\epsilon)\,,  \quad \forall t > 0\,, \label{eq:main-bound} 
    \end{align}
    and 
    \begin{align}
        |X(t)| \leq& \beta_2^\ast(\Upsilon(0), t) + \alpha_8^\ast(\overline{\Delta D}) + \alpha_9^\ast(\epsilon)\,, \label{eq:x-bound-main} \\
        \|u[t]\|_{L^1}[0, D] \leq& \beta_2^\ast(\Upsilon(0), t)  + \alpha_8^\ast(\overline{\Delta D}) + \alpha_9^\ast(\epsilon)  \,. \label{eq:u-bound-main}
    \end{align}
\end{theorem}

Theorem \ref{thm:main-result-no-measurement} is inherently a local result in that it requires the set of initial condition radius to be smaller then $R_\Upsilon$ which inherently decreases as the approximation error worsens. Hence, only for $\epsilon < \epsilon^\ast$ does a local region of attraction exist, but qualitatively, the result ensures that an $\epsilon^\ast$ always exists - although it can be extremely small and inherently unachievable in practice. In the case of which $\epsilon = 0$, we recover the local result of \cite[Theorem 3]{delphine}. Furthermore, it may seem odd that the analysis and hence the result is characterized in the $L^1$ norm; however, this is expected as one would require additional assumptions for the $L^2$ norm (c.f. \cite{delphine}).

\begin{pf}
    First, we begin by characterizing the PDE describing $\hat{u}$ as well as the corresponding target system. By direct calculation, we have
    \begin{align}
        \frac{\partial \hat{u}(x, t)}{\partial x} &= U'(t+\hat{D}(t)(x-1)) \hat{D}(t)\,, \\
        \frac{\partial \hat{u}(x, t)}{\partial t} &= U'(t+\hat{D}(t)(x-1))(1+\dot{\hat{D}}(t)(x-1))\,.
    \end{align}
    Therefore, $\hat{u}(x, t)$ satisfies the PDE
    \begin{subequations}
    \begin{align}
        \hat{D}(t) \hat{u}_t(x, t) &= \hat{u}_x(x, t) + \dot{\hat{D}}(t) (x-1) \hat{u}_x(x, t)\,, \label{eq:transport-adaptive-1}\\
        \hat{u}(1, t) &= U(t)\,. \label{eq:transport-adaptive-2}
    \end{align}
    \end{subequations}
    Thus, under the backstepping transform with the exact predictor, we obtain the following target system:
    \begin{lemma} \label{lemma:backstepping-lemma-distributed-input}
        Under the backstepping transformation for the distributed system \eqref{eq:transport-adaptive-1}, \eqref{eq:transport-adaptive-2} given by 
        \begin{align}
            \hat{w}(x, t) = \hat{u}(x, t) - \kappa(\hat{p}(x, t)),
        \end{align} the plant \eqref{eq:main-problem} becomes 
        \begin{subequations}
        \begin{align}
            \dot{X}(t) =& f(X(t), \kappa(X(t)) + \hat{w}(0, t) + \hat{u}(0, t)), \label{eq:distributed-what-1} \\
    \hat{D}(t) \hat{w}_t(x, t) = &\hat{w}_x(x, t) + \dot{\hat{D}}(t) g_3(x, t) \nonumber \\ & - g_4(x , t) f_{\tilde{u}}(t)\label{eq:distributed-what-2} \\ 
    \hat{w}(1, t)  =& \kappa(\hat{p}_{\text{NO}}(1, t) ) - \kappa(\hat{p}(1, t))\,,\label{eq:distributed-what-3}  \\ 
        D \tilde{u}_t(x, t) =& \tilde{u}_x(x, t) - \tilde{D}(t) g_1(x, t) \nonumber \\ &- \dot{\hat{D}}(t) g_2(x, t)\,, \label{eq:distributed-u-error-1} \\
    \tilde{u}(1, t) =& 0\label{eq:distributed-u-error-2} 
\end{align} 
\end{subequations}
where $\tilde{u} = u - \hat{u}$ represents the difference between the true actuation and the estimate. 
\begin{align}
  g_1(x, t) \coloneq& \frac{1 }{\hat{D}(t)}  \bigg[\hat{w}_x(x, t) + \hat{D}(t) \frac{d \kappa(\hat{p}(x, t))}{d \hat{p}} \nonumber \\ &\times f(\hat{p}(x, t), \hat{w}(x, t) + \kappa(\hat{p}(x, t)))\bigg]\,, \label{eq:g1-def} \\
  g_2(x, t) \coloneq& \frac{D(x-1)}{\hat{D}(t)} \bigg[\hat{w}_x(x, t) + \hat{D}(t) \frac{d \kappa(\hat{p}(x, t))}{d \hat{p}}\nonumber  \\ & \times f(\hat{p}(x, t), \hat{w}(x, t) + \kappa(\hat{p}(x, t)))\bigg]\,, \label{eq:g2-def}\\ 
    g_3(x, t) \coloneq& (x-1) \bigg[ \hat{w}_x(x, t) + \kappa'(\hat{p}(x, t)) \hat{D}(t) \nonumber  \\ & \times f(\hat{p}(x, t), \hat{w}(x, t) + \kappa(\hat{p}(x, t)))\bigg] \nonumber \\ 
    &- \kappa'(\hat{p}(x, t)) \hat{D}(t) \int_0^x \hat{\Phi}(x, y, t) \nonumber \\ & \times \bigg[f(\hat{p}(y, t), \hat{w}(y, t) + \kappa(\hat{p}(y, t))) \nonumber  \\ 
    & + \frac{\partial f(\hat{p}(y, t), \hat{w}(y, t) + \kappa(\hat{p}(y, t))}{\partial \hat{u}}(y-1) \nonumber \\ 
    & \times \bigg(\hat{w}_x(y, t) + \kappa'(\hat{p}(y, t))\hat{D}(t) f(\hat{p}(y, t), \hat{w}(y, t) \nonumber \\ &+ \kappa(\hat{p}(y, t))) \bigg)\bigg]dy\label{eq:g3-def} \\
    g_4(x, t) \coloneq& \hat{\Phi}(x, 0, t) \hat{D}(t) \kappa'(\hat{p}(x, t)) \,, \\
    f_{\tilde{u}}(t) \coloneq& f(\hat{p}(0, t), u(0, t)) - f(\hat{p}(0, t), \hat{u}(0, t))\,,\label{eq:g4-def}
\end{align}
and $\hat{\Phi}$ is the state transition matrix associated with the spatially-varying differential equation governing $\hat{r}(x) = \hat{D}(t)\hat{p}_t(x, t) - \hat{p}_x(x, t)$. 
    \end{lemma}
    \begin{pf}
        See Appendix \ref{appendix:proof-lemma-4}.
    \end{pf}
Note, as in Lemma \ref{lemma:backstepping-lemma}, the challenge of the target system is the non-vanishing backstepping error due to the neural operator in \eqref{eq:distributed-what-3}. 
Furthermore, by direct calculation of \eqref{eq:distributed-u-error-1}, \eqref{eq:distributed-u-error-2}, \eqref{eq:distributed-what-1}, \eqref{eq:distributed-what-2}, we obtain
\begin{align}
    D \tilde{u}_{xt}(x, t) =& \tilde{u}_{xx}(x, t) - \tilde{D}(t) g_5(x, t) \nonumber \\ & - \dot{\hat{D}}(t) g_6(x, t)\,, \label{eq:uxt-1} \\\label{eq:uxt-2}
    \tilde{u}_x(1, t) =& \tilde{D}(t)g_1(1, t)\,,  \\\label{eq:wxt-1}
    \hat{D}(t)\hat{w}_{xt}(x, t) =& \hat{w}_{xx}(x, t) + \dot{\hat{D}}(t) g_7(x, t) \nonumber \\ &- g_8(x, t) f_{\tilde{u}}(t)\,, \\ \label{eq:wxt-2}
    \hat{w}_x(1, t) =& -\dot{\hat{D}}(t)g_3(1, t) + g_4(1, t) f_{\tilde{u}}(t)\,,  \\ \label{eq:wxxt-1}
    \hat{D}(t) \hat{w}_{xxt}(x, t) &= \hat{w}_{xxx}(x, t) + \dot{\hat{D}}(t) g_9(x, t) \nonumber \\ &- g_{10}(x, t) f_{\tilde{u}}(t)\,, \\ \label{eq:wxxt-2}
    \hat{w}_{xx}(1, t) =& -\dot{\hat{D}}(t)g_7(1, t) +g_8(1, t)f_{\tilde{u}}(t) \nonumber \\ &+ \hat{D}(t)\hat{w}_{xt}(1, t)\,,
\end{align}
 where $g_5 = g_{1, x}$, $g_6 = g_{2, x}$, $q_7=g_{3, x}$, $q_8=q_{4, x}$, $q_9 = q_{7, x}$, and $g_{10} = g_{8, x}$. We refer the reader to \ref{appendix:additional-expressions} for the full expressions and omit them for brevity here. 

Consider now the Lyapunov-Krasovskii candidate 
\begin{align}
    \hat{W}(t) =& V_0(X) + b_0D\int_0^1 (1+x) |\tilde{u}(x, t)| dx \nonumber \\ 
    &+ b_1 D \int_0^1 (1+x) |\tilde{u}_x(x, t)| dx \nonumber \\ &+ b_2 \hat{D}(t) \int_0^1 (1+x)|\hat{w}(x, t)| dx \nonumber \\ 
    &+ b_3 \hat{D}(t) \int_0^1 (1+x)|\hat{w}_x(x, t)| dx \nonumber \\ 
    &+ b_4 \hat{D}(t) \int_0^1 (1+x) |\hat{w}_{xx}(x, t) |dx + \tilde{D}(t)^2 \,,
\end{align}
where $V_0 = \sqrt{V}$ as in Assumption \ref{assumption:gas} and $b_0, b_1, b_2, b_3, b_4 > 0$ are constants to be specified.

First, by the mean value theorem, we have that 
\begin{align}
    \dot{V}_0 =& \frac{1}{2} \frac{dV_0}{dX}(f(X, \kappa(X)+\tilde{u}(0, t) + \hat{w}(0, t)) \nonumber  \\ 
  %   =& \frac{1}{2}\frac{dV_0}{dX}(f(X, \kappa(X))) +  \frac{1}{2}\frac{dV_0}{dX} (f(X, \kappa(X)+\tilde{u}(0, t) + \hat{w}(0, t)) - f(X, \kappa(X)) \nonumber \\ 
     \leq& -\frac{\lambda}{2}|X| + \frac{C_2}{2} \left| \frac{\partial f(X, a_1(X, t) )}{\partial u} \right| \left| \hat{w}(0, t) + \tilde{u}(0, t) \right|
\end{align}
where $a_1(X, t) \in [\kappa(X(t)), \kappa(X(t)) + \hat{u}(0, t)]$. Substituting and applying integration by parts yields
\begin{align}
        \dot{\hat{W}}(t) \leq &  -\lambda/2 |X| + \frac{C_2}{2} \left|\frac{\partial f(x, \alpha_1(x, t))}{\partial u} \right| \left| \hat{w}(0, t) + \tilde{u}(0, t) \right| \nonumber 
        \\ & + b_0 \bigg[ - |\tilde{u}(0, t)|  - \|\tilde{u}[t]\|_{L^1} \nonumber \\ & + |\tilde{D}(t)| \int_0^1 (1+x)|g_1(x, t)|dx  \nonumber \\ & + |\dot{\hat{D}}(t)| \int_0^1 (1+x) |g_2(x, t)| dx   \bigg] \nonumber
        \\ & + b_1 \bigg[2|\tilde{u}_x(1, t)|  - |\tilde{u}_x(0, t)| - \|\tilde{u}_x[t]\|_{L^1} \nonumber \\ & +  |\tilde{D}(t)|\int_0^1 (1+x) |g_5(x, t)| dx \nonumber \\ & + |\dot{\hat{D}}(t)|\int_0^1 (1+x) |g_6(x, t)|dx \nonumber \bigg]  
        \\ & \nonumber + b_2 \bigg[ 2|\hat{w}(1, t)| - |\hat{w}(0, t)| - \|\hat{w}[t]\|_{L^1} \nonumber \\ & + |\dot{\hat{D}}(t)| \int_0^1 (1+x) |g_3(x, t)| dx  \nonumber \\ & |f_{\tilde{u}}(t)| \int_0^1 (1+x) |g_4(x, t)|dx \nonumber \bigg]
        \\ & \nonumber + b_3 \bigg[ 2|\hat{w}_x(1, t)| - |\hat{w}_x(0, t)|  - \|\hat{w}_x[t]\|_{L^1}  \nonumber \\ &  + |\dot{\hat{D}}(t)|\int_0^1 (1+x)|g_7(x, t)|dx \nonumber \\ & + |f_{\tilde{u}}(t)| \int_0^1 (1+x)| g_8(x, t)| dx  \bigg] \nonumber
        \\ & \nonumber  +b_4 \bigg[2|\hat{w}_{xx}(1, t)| - |\hat{w}_{xx}(0, t)|  - \|\hat{w}_{xx}[t]\|_{L^1} \nonumber \\ &  + |\dot{\hat{D}}(t)|\int_0^1 (1+x)|g_9(x, t)|dx \nonumber \\ & + |f_{\tilde{u}}(t)| \int_0^1 (1+x) |g_{10}(x, t)|dx \bigg]  \nonumber \\ &  + 2 \tilde{D}(t) \dot{\hat{D}}(t) \nonumber \\ 
    &+ b_2 \dot{\hat{D}}(t) \int_0^1 (1+x)|\hat{w}(x, t)| dx \nonumber \\ 
      &+ b_3 \dot{\hat{D}}(t) \int_0^1 (1+x)|\hat{w}_x(x, t)| dx  \nonumber \\ 
        &+ b_4 \dot{\hat{D}}(t) \int_0^1 (1+x)|\hat{w}_{xx}(x, t)| dx 
\end{align}

Now, define the following functional
\begin{align}
    \hat{W}_0(t) =& |X(t)| + \|\tilde{u}(t)\|_{L^1} + \|\tilde{u}_x(t)\|_{L^1} +\|\hat{w}(t)\|_{L^1} \nonumber \\&+\|\hat{w}_x(t)\|_{L^1} + \|\hat{w}_{xx}(t)\|_{L^1}\,. 
\end{align}
 Note, there exist $\alpha_{10}, \alpha_{11} \in \mathcal{K}_\infty$ such that 
\begin{align}
    \frac{C_2}{2}\left| \frac{\partial f}{\partial u}(X, \alpha_1(X, t)) \right| \leq \alpha_{10}( \hat{W}_0(t)) + \alpha_{11}(\epsilon)\,.
\end{align}
Furthermore, let the constant $\eta = \min\{\lambda/2, b_0, b_1, b_2, b_3, b_4 \}$ for brevity. Applying Lemmas \ref{lemma:g1-g2}, \ref{lemma:g3-g4}, \ref{lemma:g5-g6}, \ref{lemma:g7-g8}, \ref{lemma:g9-g10}, (See Appendix \ref{appendix:proof-tech-expressions}) for brevity. Applying Lemmas 7, 8, 9, 10, 11, and usingand using the properties of class $\mathcal{K}_\infty$ functions results in 
\begin{align}
\dot{\hat{W}}(t) \leq& \; - \eta \hat{W}_0(t) \nonumber \\ 
& + |\dot{\hat{D}}(t)| \bigg[ b_0 \alpha_{17}(\hat{W}_0) 
+ b_1 \big(2\alpha_{24}(\hat{W}_0) + 2\alpha_{24}(\epsilon)\big) \nonumber\\ 
& + b_2 \big(2\alpha_{19}(\hat{W}_0) + 2\alpha_{19}(\epsilon)\big) \nonumber \\ 
& + 2b_3 \big(2\alpha_{21}(\hat{W}_0) + 2\alpha_{21}(\epsilon)\big) \nonumber \\ 
& + b_3 \big(2\alpha_{25}(\hat{W}_0) + 2\alpha_{25}(\epsilon)\big) \nonumber \\  
& + b_4 \big(2\alpha_{27}(\hat{W}_0) + 2\alpha_{27}(\epsilon)\big) \nonumber \\ 
& + 2(b_2+b_3+b_4)\hat{W}_0(t) \bigg] \nonumber \\ 
& + 2b_4 \big(|\dot{\hat{D}}(t)| + |\dot{\hat{D}}(t)|^2 + |\dot{\hat{D}}(t)|^3 \big) \nonumber \\ 
& \big(2\alpha_{29}(\hat{W}_0) + 2\alpha_{29}(\epsilon)\big) \nonumber \\ 
& + 2b_4 |\ddot{D}(t)| \big(2\alpha_{33}(\hat{W}_0) + 2\alpha_{33}(\epsilon)\big) \nonumber \\ 
& + |\tilde{D}(t)| \bigg[ b_0 \alpha_{16}(\hat{W}_0) 
+ 2b_1 M_4 M_1 M_3 \epsilon + 2b_1 \alpha_{18}(\hat{W}_0) \nonumber \\ 
& + b_1 \big(2\alpha_{23}(\hat{W}_0) + 2\alpha_{23}(\epsilon)\big)  \nonumber \\ 
& + 2b_4 \big(2\alpha_{34}(\hat{W}_0) + 2\alpha_{34}(\epsilon)\big) \bigg] \nonumber \\ 
& + |\tilde{u}_x(0, t)| \bigg[ -b_1 
+ 2b_4 \big(2\alpha_{30}(\hat{W}_0) + 2\alpha_{30}(\epsilon)\big) \bigg] \nonumber \\ 
& + |\hat{w}_x(0, t)| \bigg[ -b_3 
+ 2b_4 \big(2\alpha_{31}(\hat{W}_0) + 2\alpha_{31}(\epsilon)\big) \bigg] \nonumber \\ 
& + |\tilde{u}(0, t)| \bigg[ -b_0 + \alpha_{10}(\hat{W}_0(t)) + \alpha_{11}(\epsilon) \nonumber \\ 
& + b_2 \big(2\alpha_{20}(\hat{W}_0) + 2\alpha_{20}(\epsilon)\big)  \nonumber \\ 
& + 2b_3 \big(2\alpha_{22}(\hat{W}_0) + 2\alpha_{22}(\epsilon)\big) \nonumber \\ 
& + b_3 \big(2\alpha_{26}(\hat{W}_0) + 2\alpha_{26}(\epsilon)\big) \nonumber \\ 
& + 2b_4 \big(2\alpha_{32}(\hat{W}_0) + 2\alpha_{32}(\epsilon)\big) \nonumber \\ 
& + b_4 \big(2\alpha_{28}(\hat{W}_0) + 2\alpha_{28}(\epsilon)\big) \bigg] \nonumber \\ 
& + |\hat{w}(0, t)| \bigg[ -b_2 + \alpha_{10}(\hat{W}_0(t)) + \alpha_{11}(\epsilon) \bigg] \nonumber \\ 
& + |\hat{w}_{xx}(0, t)| \, b_4 \bigg[-1 + |\dot{\hat{D}}(t)| \bigg] \nonumber \\ 
& + 2 b_2 M_2 \epsilon \nonumber \\ 
& + 2 b_1 |\tilde{D}(t)| \bigg[ M_5 \Big( |\dot{\hat{D}}(t)| (2\alpha_{21}(\hat{W}_0) + 2\alpha_{21}(\epsilon)) \nonumber \\ 
& +|\tilde{u}(0, t)| (2\alpha_{22}(\hat{W}_0) + 2\alpha_{22}(\epsilon)) \Big) \bigg]
\end{align}
    To handle the cross terms, we use the following two facts:
    \begin{align}
        |\tilde{D}(t)|\leq& 2\overline{D} =: M_6\,, \\
        |\dot{\hat{D}}(t)| \leq & \gamma \alpha_{1, D} (\Gamma_0(t)) \leq \gamma \alpha_{35}(\hat{W}_0(t))\,, 
    \end{align}
    where $\alpha_{35} \in \mathcal{K}_\infty$. 
Further, the parameters can cancel the boundaries within a local initial region of attraction. Let $R = \hat{W}_0(0)$. Then, given the conditions below hold:    \begin{align}
        b_1>& 2b_4(2\alpha_{30}(R)+2\alpha_{30}(\epsilon))\,, \\ 
        b_3>& 2b_4(2\alpha_{31}(R)+2\alpha_{31}(\epsilon))\,, \\
        b_2>& \alpha_{10}(R)+\alpha_{11}(\epsilon)\,, \\
        b_0>&  \alpha_{10}(R)+\alpha_{11}(\epsilon) + b_2(2\alpha_{20}(R)+2\alpha_{20}(\epsilon)) \nonumber \\ &  +2b_3 (2\alpha_{22}(R)+2\alpha_{22}(\epsilon)) + b_3(2\alpha_{26}(R)+2\alpha_{26}(\epsilon)) \nonumber \\ & +2b_4 (2\alpha_{32}(R)+2\alpha_{32}(\epsilon))+b_4(2\alpha_{28}(R)+2\alpha_{28}(\epsilon))\nonumber \\ &
        2b_1M_5M_6(2\alpha_{22}(R)+2\alpha(\epsilon))\,,
    \end{align}
    we obtain
\begin{align}
\dot{\hat{W}}(t) \leq& - \eta \hat{W}_0(t) 
+ |\dot{\hat{D}}(t)|\!\big[ b_0\alpha_{17}(\hat{W}_0) \nonumber\\ 
& + b_1(2\alpha_{24}(\hat{W}_0)+2\alpha_{24}(\epsilon))  \nonumber \\ 
&  + b_2(2\alpha_{19}(\hat{W}_0)+2\alpha_{19}(\epsilon))  \nonumber \\ 
& + 2b_3(2\alpha_{21}(\hat{W}_0)+2\alpha_{21}(\epsilon))  \nonumber \\ 
&  + b_3(2\alpha_{25}(\hat{W}_0)+2\alpha_{25}(\epsilon)) \nonumber \\ 
& + b_4(2\alpha_{27}(\hat{W}_0)+2\alpha_{27}(\epsilon)) \nonumber \\ 
& + 2(b_2+b_3+b_4)\hat{W}_0(t) \big] \nonumber \\ 
& + 2b_4(|\dot{\hat{D}}(t)|+|\dot{\hat{D}}(t)|^2+|\dot{\hat{D}}(t)|^3) \nonumber \\ 
& \times (2\alpha_{29}(\hat{W}_0)+2\alpha_{29}(\epsilon)) \nonumber \\ 
& + 2b_4|\ddot{D}(t)|(2\alpha_{33}(\hat{W}_0)+2\alpha_{33}(\epsilon)) \nonumber \\ 
& + |\tilde{D}(t)|\!\big[ b_0\alpha_{16}(\hat{W}_0)+2b_1M_4M_1M_3\epsilon  \nonumber \\ 
& + 2b_1\alpha_{18}(\hat{W}_0)  \nonumber \\ 
&  + b_1(2\alpha_{23}(\hat{W}_0)+2\alpha_{23}(\epsilon)) \nonumber \\ 
& + 2b_4(2\alpha_{34}(\hat{W}_0)+2\alpha_{34}(\epsilon))  \nonumber \\ 
&  + 2b_1M_5\gamma \alpha_{35}(\hat{W}_0)(2\alpha_{21}(\hat{W}_0)+2\alpha_{21}(\epsilon)) \big] \nonumber \\ 
& + |\hat{w}_{xx}(0,t)|b_4(-1+|\dot{\hat{D}}(t)|) + 2b_2M_2\epsilon 
\end{align}

     Now, let us define the functions
    \begin{align}
        \alpha_{36}(\hat{W}_0(t)) =&  b_0\alpha_{17}(\hat{W}_0(t))+ b_12\alpha_{24}(\hat{W}_0(t)) \nonumber\\ & + b_22\alpha_{19}(\hat{W}_0(t)) + 4b_3 \alpha_{21}(\hat{W}_0(t)) \nonumber \\ & + b_3 2\alpha_{25}(\hat{W}_0(t))   + b_4 2\alpha_{27}(\hat{W}_0(t))  \nonumber \\ & + 2(b_2+b_3+b_4)\hat{W}_0(t)\,, \\  
        \alpha_{37}(\epsilon) =& b_12\alpha_{24}(\epsilon) \nonumber\\ & + b_22\alpha_{19}(\epsilon) + 4b_3 \alpha_{21}(\epsilon) \nonumber \\ & + b_3 2\alpha_{25}(\epsilon)   + b_4 2\alpha_{27}(\epsilon) \,,\\
        \alpha_{38}(\hat{W}_0(t)) =& b_0\alpha_{16}(\hat{W}_0(t)) \nonumber \\ & + 2b_1\alpha_{18}(\hat{W}_0(t)) + 2b_1\alpha_{23}(\hat{W}_0(t)) \nonumber \\ & +4b_4 \alpha_{34}(\hat{W}_0(t)) \nonumber\\ &  + 4b_1M_5\gamma \alpha_{35}(\hat{W}_0(t))\alpha_{21}(\hat{W}_0(t))\,, \\
        \alpha_{39}(\epsilon) =& 2b_1M_4M_1M_3\epsilon + 2b_1\alpha_{23}(\epsilon) + 4b_4\alpha_{34}(\epsilon) \,. 
    \end{align}
Using Assumption \ref{assumption:delay-func}, and the class $\mathcal{K}_\infty$ functions relating $\Xi$ to $\hat{W}_0(t)$, we have $|\dot{\hat{D}}(t)| \leq \gamma \alpha_{40}(\hat{W}_0(t)) $ and $|\ddot{\hat{D}}(t)|\leq \gamma \alpha_{41}(\hat{W}_0(t))$. Let $\gamma \in [0, 1]$, then we obtain    \begin{align}
        \dot{\hat{W}}(t) \leq& -\eta \hat{W}_0(t) +\gamma \alpha_{40}(\hat{W}_0(t)) (\alpha_{36}(\hat{W}_0(t)) + \alpha_{37}(\epsilon)) \nonumber \\ 
           &+ 2b_4\gamma (\alpha_{40}(\hat{W}_0(t)) + \alpha_{40}^2(\hat{W}_0(t))+\alpha_{40}^3 (\hat{W}_0(t))) \nonumber\\ 
           & \times (2\alpha_{29}(\hat{W}_0(t)) + 2 \alpha_{29}(\epsilon)) \nonumber \\ & + 2b_4\alpha_{41}(\hat{W}_0(t))(2\alpha_{33}(\hat{W}_0(t)) + 2 \alpha_{33}(\epsilon)) \nonumber \\ & +|\tilde{D}(t)|(\alpha_{38}(\hat{W}_0(t)) + \alpha_{39}(\epsilon) \nonumber \\ & + 4b_1M_5\gamma \alpha_{35}(\hat{W}_0(t)) \alpha_{21}(\epsilon)) \nonumber \\ & 
            +|\hat{w}_{xx}(0, t)|b_4\bigg[-1 + \gamma\alpha_{40}(\hat{W}_0(t))\bigg] + 2b_2M_2 \epsilon \,. 
    \end{align}
       Via Young's inequality, there exists $\alpha_{43:46} \in \mathcal{K}_\infty$ where
    \begin{align}
        \dot{\hat{W}}(t) \leq& -\eta \hat{W}_0(t) + \gamma(\alpha_{43}(\hat{W}_0(t)) + \alpha_{44}(\epsilon)) \nonumber \\ 
        & +|\tilde{D}(t)|(\alpha_{45}(\hat{W}_0(t)) + \alpha_{46}(\epsilon)) \nonumber+ 2b_2M_2 \epsilon \\ & 
            +|\hat{w}_{xx}(0, t)|b_4\bigg[-1 +  \gamma\alpha_{40}(\hat{W}_0(t))\bigg]  \,. 
    \end{align}
    Using Young's inequality, for any $\delta > 0$ small, we have
    \begin{align}
        |\tilde{D}(t)| \leq \frac{\delta}{2} + \frac{1}{2\delta}|\tilde{D}^2(t)| \leq \frac{\delta}{2} + \frac{1}{2\delta }\hat{W}(t)\,. 
    \end{align}
    Hence, we obtain 
      \begin{align}
        \dot{\hat{W}}(t) \leq& -\eta \hat{W}_0(t) + \gamma(\alpha_{42}(\hat{W}_0(t)) + \alpha_{43}(\epsilon)) \nonumber \\ 
        & +(\frac{\delta}{2} + \frac{1}{2\delta }\hat{W}(t))(\alpha_{44}(\hat{W}_0(t)) + \alpha_{45}(\epsilon)) \nonumber \\ & 
            +|\hat{w}_{xx}(0, t)|b_4\bigg[-1 + \gamma\alpha_{40}(\hat{W}_0(t))\bigg] + 2b_2M_2 \epsilon \,. 
    \end{align}
    Now, let $\nu \in (0, 1)$ and define $C_{R_1} := \max_{x \in [0, R]} \alpha'_{42}(x)$, $C_{R_2} := \max_{x \in [0, R]}\alpha'_{44}(x) $. Then, there exists $\alpha_{46}, \alpha_{47} \in \mathcal{K}_\infty$ such that if  
    \begin{align}
        \gamma <& \gamma^\ast := \min \bigg\{ 1, \frac{1}{\max_{x \in [0, R]}\alpha_{40}(x)}, \frac{\eta \nu}{C_{R_1}} \bigg\}\,, \\ 
        \delta <& 2 \frac{\eta \nu - \gamma C_{R_1}}{C_{R_2}} \,, \\
 \hat{W}(0) < & 2 \delta \frac{\eta \nu -\gamma C_{R_1}-\delta/2 C_{R_2}-\alpha_{46}(\epsilon)}{C_{R_2}+\alpha_{45}(\epsilon)} \,, 
    \end{align}
    then we have
    \begin{align}
        \dot{\hat{W}}(t) \leq -\eta(1-\nu)\hat{W}_0(t)  + \alpha_{47}(\epsilon)\,. 
    \end{align}
    Note that $\tilde{D}^2$ is bounded by $(2\overline{\Delta D})^2$. Further,  $|X| \leq V_0(X) \leq \sqrt{C_1}|X|$ by \eqref{eq:linear-lyapunov-2}. Thus, $\hat{W}_0(t)$ and $\hat{W}$ are equivalent up to $\tilde{D}^2$ such that there exists constants $C_3, C_4>0$
    \begin{align}
        C_3 \hat{W}_0(t) \leq \hat{W}(t) \leq C_4(\hat{W}_0(t) + \tilde{D}^2(t))\,. 
    \end{align}
    Hence, there exists $\mathcal{K}_\infty$ functions $\alpha_{48}, \alpha_{49}, \alpha_{50}$ such that 
    \begin{align}
        \dot{\hat{W}}(t) \leq -\alpha_{49}(\hat{W}(t)) + \alpha_{50}(\overline{\Delta D}) + \alpha_{51}(\epsilon)\,. 
    \end{align}
    From here, using \cite[Lemma C.3]{kkk}, we have there exists $\beta_3 \in \mathcal{KL}, \alpha_{52}, \alpha_{53} \in \mathcal{K}_\infty$ such that
    \begin{align}
        \hat{W}(t) \leq \beta_3(\hat{W}(0), t) + \alpha_{52}(\overline{\Delta D}) + \alpha_{53}(\epsilon)\,.
    \end{align}
    To conclude the result, notice that $\Upsilon$ and $\hat{W}$ are equivalent up to class $\mathcal{K}_\infty$ functions to obtain \eqref{eq:main-bound}. To obtain the bound on $|X(t)|$, notice that 
    \begin{align}|X(t)| &\leq V_0(t) \nonumber  \\ &< W(t) \nonumber  \\ & \leq \beta_3(\hat{W}(0), t) + \alpha_{52}(\overline{\Delta D}) + \alpha_{53}(\epsilon) \,,
    \end{align}
    and using the equivalence again between $\Upsilon$ and $\hat{W}$ yields the result. The exact same approach can be used to obtain the bound on $\|u\|_{L^1}$. 
    
% Applying the equivalence of class $K_\infty$ functions from $\hat{W}$ to $\Upsilon $ along with the bound on $W(0)$ yields the final result. Regulation, is shown in the exact same manner as in Section \ref{sec:stability} using the relation of class $\mathcal{K}$ functions between $\hat{W}$ and $X$ and $U$. 

\end{pf}

\section{Numerical Example: Biological Chemostat} \label{sec:chemostat}
To illustrate neural operator predictors with unknown actuator input, we consider a Chemostat—a bioreactor for growing micro-organisms or cells under continuous, controlled conditions \cite{Smith_Waltman_1995}. Chemostats are widely used in biological engineering, from wastewater treatment \cite{1099745} to genetic engineering via recombinant DNA \cite{STEPHANOPOULOS198849,Levin1980-wm}. Here, we focus on a recent Chemostat model that incorporates population mortality into the dynamics \cite{karafyllis2025globalstabilizationchemostatsnonzero}. Consider the nonlinear system
\begin{subequations} \label{eq:chemostat}
    \begin{align}
        \dot{Z}(t) &= (\rho_0 \mu(S) - \chi- U(t-D))Z(t)\,, \\
        \dot{S}(t) &= U(t-D)(S_{\text{in}}-S) - \mu(S)Z(t)\,,
    \end{align}
\end{subequations}
where $Z(t)>0$ is the microbial concentration, $S(t)>0$ the substrate concentration, $S_\text{in}>0$ the inlet substrate concentration, $U(t)>0$ the dilution rate, $\rho_0>0$ the yield coefficient, $\mu$ the population growth rate, $\chi\ge 0$ the mortality rate, and $D$ the dilution delay (i.e. due to valve transport or homogenization time). Without delay, a nominal globally stabilizing controller is given by \cite{karafyllis2025globalstabilizationchemostatsnonzero}
\begin{align}
    \kappa(Z(t)&, S(t)) \nonumber \\ =&  \frac{U^\ast \mu(S) Z(t)}{\mu(S^\ast)Z^\ast} \nonumber \\ &+  \frac{\varsigma \cdot \chi}{(\mu(S^\ast))^{1+\xi}} \begin{cases}
    |\mu(S) - \mu(S^\ast)|^{1+\xi}\,, &\text{if } S\leq S^\ast\\
    0 &\text{if } S> S^\ast \,. 
    \end{cases}
 \end{align}
 where  $S^\ast$ is the desired substrate concentration, $Z^\ast$ is the desired microbial population concentration, $U^\ast$ is the desired dilution rate and $\varsigma > 0$ and $\alpha \in [0, 1)$ are constant gains chosen by the user. 
 
 In the experiments that follow, we set the following parameters:  $Z^\ast = 3$, $S^\ast = 2$, $U^\ast = 0.9$, $\varsigma=10$, $\chi=0.1$, $S_{\text{in}} = 5.33$, $\alpha=0.5$, $\rho_0 = 1$ and the growth rate function
 \begin{align}
     \mu(S) = \frac{7S}{2(1+S+S^2)}\,, 
 \end{align}
where we note $(Z^\ast, S^\ast)$ is an unstable equilibrium. 

\begin{figure*}[htbp]
 \centering 
     \includegraphics[width=\textwidth]{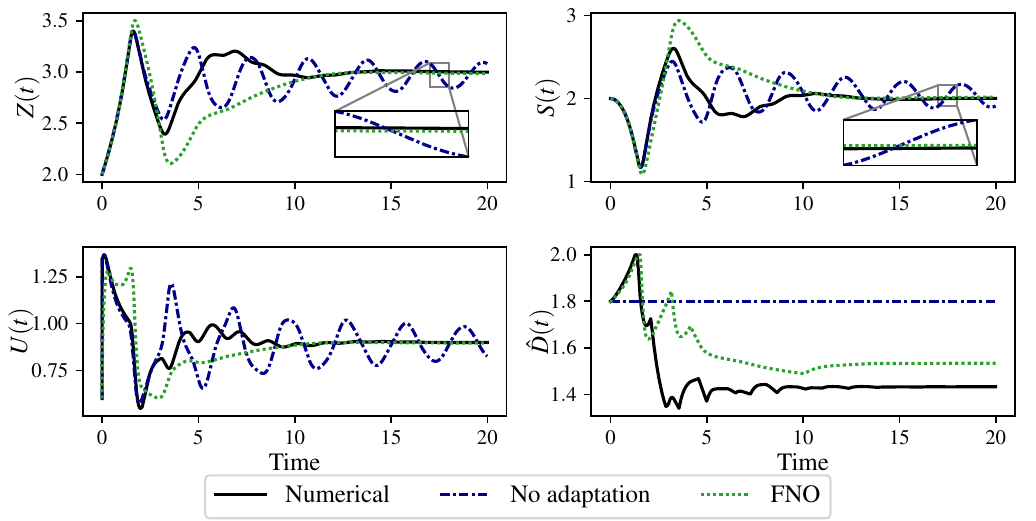}
    \caption{Chemostat \eqref{eq:chemostat} with $Z(0)=2,S(0)=2$, $\hat{D}(0)=1.8$, with target setpoint at the unstable equilibrium $Z^\ast=3,S^\ast=2$ \cite{karafyllis2025globalstabilizationchemostatsnonzero}. For true delay $D=1.6$ s, comparison of (i) numerical predictor (black), (ii) open-loop without adaptation, unstable (blue), and (iii) FNO predictor, stable (green).}
    \label{fig:chemostat-main}
 \end{figure*}

As in Section \ref{sec:numerical}, we generated a dataset of $100,000$ predictor input–output pairs for training by uniformly varying initial conditions and delay values. Consistent with theory, the unknown actuated input required a smaller $\epsilon$ tolerance, and thus more training data and a larger network to achieve stabilization. Among tested architectures, only the FNO stabilized the system, achieving an order-of-magnitude lower test loss; we expect DeepONet could perform similarly with extensive tuning \cite{LU2022114778}. Figure \ref{fig:chemostat-main} shows that, without delay compensation, the system oscillates to a limit cycle. However, with compensation, the FNO predictor stabilizes the system. Moreover, Table \ref{tab:comp-chemostat} reports computation costs, with the FNO giving an $8\times$ speedup over the numerical approach at the smallest discretization.
% \usepackage{graphicx}
\begin{table}[htpb]
\centering
\caption{Computation time (ms) for various approximate predictors averaged over 1000 samples for the Chemostat example.}
\label{tab:comp-chemostat}
\resizebox{0.47\textwidth}{!}{%
\begin{tabular}{l|lllll}
\begin{tabular}[c]{@{}l@{}}Step \\ size (dx)\end{tabular} & Numerical $\downarrow$ & DeepONet $\downarrow$ & FNO $\downarrow$ & \begin{tabular}[c]{@{}l@{}}DeepONet\\ Speedup $\uparrow$\end{tabular} & \begin{tabular}[c]{@{}l@{}}FNO\\ Speedup $\uparrow$\end{tabular} \\ \hline
$0.01$                                                    & $1.40$                 & $\bm{0.56}$                & $1.28$           & $\bm{2.50\times}$                                                          & $1.09\times$                                                     \\
$0.005$                                                   & $2.86$                 & $\bm{0.53}$                & $1.34$           & $\bm{5.43\times}$                                                          & $2.13\times$                                                     \\
$0.001$                                                   & $15.62$                & $\bm{1.01}$                & $1.88$           & $\bm{15.48\times}$                                                         & $8.29\times$                                                    
\end{tabular}%
}
\end{table}

 % \begin{figure*}
 %    \label{fig:predictors-main}
 %     \includegraphics{figures/resErrors.pdf}
 %    \caption{For the same simulations as in Figure \ref{fig:chemostat-main}, above is the corresponding approximate predictor values for each approach as well as the predictor errors over time.}
 % \end{figure*}

 % \begin{figure*}
 %        \label{fig:predictors-generalization}
 %     \includegraphics{figures/resGeneralization.pdf}
 %             \caption{Presented is simulations of the Chemostat \eqref{eq:chemostat}  with various initial conditions for the neural operator predictor (FNO) without retraining. The simulation's delay estimate is initialized to $\hat{D}(0) = 1.8$ and the dynamical systems contains a true delay of $D = 1.5$s. }

 % \end{figure*}

 \section{Conclusion}
We presented the first results on approximate predictors for nonlinear systems with unknown delays, considering cases with measured and unmeasured actuated inputs. In the first setting we established semi-global practical stability, while in the second we obtained local practical stability due to the underlying local result. The analysis applies to any uniform predictor, but simulations with neural operators demonstrated stability and significant computational gains over numerical methods. Case studies on an E. coli biological clock and a chemostat illustrated stability and generalization of the neural-operator approach. This work opens directions toward trajectory tracking and output-feedback designs with unknown parameters.
\bibliographystyle{elsarticle-num}
\bibliography{references}

\appendix 

\section{Reverse Young's Inequality}\label{appendix:reverse-youngs}
\begin{lemma}
    Given $a, b \geq 0$ and $p, q$ such that 
    \begin{align}
        \frac{1}{p} - \frac{1}{q} = 1\,, 
    \end{align}
    then
    \begin{align}
        \frac{a^p}{p} - \frac{b^{-q}}{q} \leq ab\,. 
    \end{align}
\end{lemma}
\begin{pf}
    See \cite{proofwiki_reverse_young}. 
\end{pf}
\section{Proof of Lemma \ref{lemma:backstepping-lemma-distributed-input}} \label{appendix:proof-lemma-4}
\begin{pf}
We omit the arguments $(x, t)$ unless necessary for space and note this Lemma is very similar to \cite[Lemma 4]{delphine}. The computation of \eqref{eq:distributed-what-1} is by direct substitution. We first compute the error system $\tilde{u}$. Direct calculation yields
\begin{align}
    \tilde{u}_x = u_x - \hat{u}_x \,, \\
    \tilde{u}_t = u_t- \hat{u}_t\,. 
\end{align}
Substituting $Du_t = u_x$ yields 
\begin{align}
    \tilde{u}_t &= \frac{1}{D} \left(\tilde{u}_x + \hat{u}_x \right)  - \hat{u}_t \nonumber  \\ 
    &=\frac{1}{D} (\tilde{u}_x + \hat{u}_x) - \frac{1+\dot{\hat{D}}(x-1)}{\hat{D}(t)} \hat{u}_x\,. 
\end{align}
Multiplying by $D$ yields
\begin{align}
    D \tilde{u}_t &= \tilde{u}_x + \left[1 - \frac{D(1+\dot{\hat{D}}(x-1))}{\hat{D}(t)}\right] \hat{u}_x \nonumber \\ 
    &=\tilde{u}_x + \left[\frac{\hat{D}(t) - D(1+(\dot{\hat{D}}(x-1)))}{\hat{D}(t)} \right] \hat{u}_x \nonumber \\ 
    &= \tilde{u}_x - \frac{\tilde{D}(t)}{\hat{D}(t)} \hat{u}_x  - \frac{D\dot{\hat{D}}(t)(x-1)}{\hat{D}(t)} \hat{u}_x\,. 
\end{align}
Substituting the following by direct calculation
\begin{align}
    \hat{w}_x = \hat{u}_x - \kappa'(\hat{p}) \hat{p}_x\,, \\
    \hat{w}_t = \hat{u}_t - \kappa'(\hat{p}) \hat{p}_t\,,
\end{align}
along with the predictor derivatives given by
\begin{align}
    \hat{p}_x =& \hat{D}(t)f(\hat{p}, \hat{u})\,, \\
    \hat{p}_t =& f(X(t), u(0, t)) + \dot{\hat{D}}(t) \int_0^x f(\hat{p}(y, t), \hat{u}(y, t)) dy \nonumber \\ &+ \hat{D}(t) \int_0^x \bigg[ \frac{\partial f(\hat{p}(y, t), \hat{u}(y, t)}{\partial \hat{p}} \times \hat{p}_t(y, t)\nonumber \\ & +  \frac{\partial f(\hat{p}(y, t), \hat{u}(y, t)}{\partial \hat{u}} \times \hat{u}_t(y, t) \bigg] dy\,, 
\end{align}
yields \eqref{eq:distributed-u-error-1} and \eqref{eq:distributed-u-error-2} is $0$ by direct substitution. To obtain \eqref{eq:distributed-what-2}, we have that multiplying $\hat{w}_t$ by $\hat{D}(t)$ yields
\begin{align}
    \hat{D}(t) \hat{w}_t = \hat{D}(t)\hat{u}_t - \hat{D}(t) \kappa'(\hat{p}) \hat{p}_t\,.
\end{align}
Substituting for $\hat{D}(t)\hat{u}_t$ yields
\begin{align}
    \hat{D}(t) \hat{w}_t =& (1+\dot{\hat{D}}(t)(x-1))\hat{u}_x - \hat{D}(t) \kappa'(\hat{p}) \hat{p}_t\,. 
\end{align}
Subtracting $\hat{w}_x$ from above yields
\begin{align}
    \hat{D}(t) \hat{w}_t - \hat{w}_x =& \dot{\hat{D}}(t)(x-1)\hat{u}_x - \kappa'(\hat{p})(\hat{D}(t) \hat{p}_t - \hat{p}_x) \label{eq:what-intermeidate}  \,. 
\end{align}
It is clear that we must obtain an expression for $\hat{D}(t)\hat{p}_t-\hat{p}_x$. Thus, we define the function $\hat{r}(x) =  \hat{D}(t)\hat{p}_t(x, t) - \hat{p}_x(x, t)$. Then, by direct computation, $\hat{r}(x)$ is governed by the following differential equation
\begin{align}
    \frac{d \hat{r}(x)}{dx} =&  \hat{D}(t) \frac{\partial f(\hat{p}(x, t), \hat{u}(x, t)}{\partial \hat{p}}\hat{r}(x) \nonumber \\ & + \hat{D}(t) \dot{\hat{D}}(t) \bigg[  f(\hat{p}(x, t), \hat{u}(x, t)) \nonumber \\ &+ \frac{\partial f(\hat{p}(x, t), \hat{u}(x, t))}{\partial \hat{u}} \times (x-1) \hat{u}_x(x, t) \bigg]\,,  \\ 
    \hat{r}(0) =& \hat{D}(t) \bigg[f(\hat{p}(0, t), u(0, t)) - f(\hat{p}(0, t), \hat{u}(0, t)) \bigg]\,. 
\end{align}
Now, allow us to define $\hat{\Phi}(x, y, t)$ as the transition matrix to the solution to this differential equation. Then, we have the relation
\begin{align}
    \hat{r}(x) =& \hat{\Phi}(x, 0, t) \hat{D}(t) \bigg[f(\hat{p}(0, t), u(0, t)) \nonumber \\ &- f(\hat{p}(0, t), \hat{u}(0, t)) \bigg]  + \dot{\hat{D}}(t)\hat{D}(t) \int_0^x \Phi(x, y, t) \nonumber \\ & \times \bigg[f(\hat{p}(y, t), \hat{u}(y, t)) \nonumber \\ & + \frac{\partial f(\hat{p}(y, t), \hat{u}(y, t))}{\partial \hat{u}}(y-1)\hat{u}_x(y, t)\bigg]dy  \label{eq:r-of-x}
\end{align}
Substituting the relation for $\hat{r}(x)$ in \eqref{eq:r-of-x} into \eqref{eq:what-intermeidate} yields \eqref{eq:distributed-what-2}. \eqref{eq:distributed-what-3} is then obtained by direct substitution of the control law in the backstepping transform. 
 
\end{pf}
\section{Expressions for $g_5, g_6, g_7, g_8, g_9, g_{10}$.} \label{appendix:additional-expressions}
We omit the arguments of the function for brevity:
\begin{align}
    g_5 =& g_{1x} \nonumber \\ 
    =&\frac{1}{\hat{D} } \bigg[\hat{w}_{xx} + \hat{D}^2 \kappa'(\hat{p})
    \times f(\hat{p}, \hat{w}+\kappa(\hat{p})) ^\top f(\hat{p}, \hat{w}+\kappa(\hat{p}))  \nonumber \\ &+ \hat{D} \kappa(\hat{p}) \times \bigg[\frac{d }{d x}f(\hat{p}(x, t), \hat{w}(x, t)+\kappa(\hat{p}(x, t))\bigg] \label{eq:g5-def}\,,  \\
    g_6 =& g_{2x} \nonumber \\ 
    =& D \times g_1  + \frac{D(x-1)}{\hat{D}}g_5 \label{eq:g6-def}\,,  \\ 
g_7 =& g_{3x}\nonumber \\ 
=&\hat{w}_x(x, t) + \kappa'(\hat{p}) \hat{D} \times f(\hat{p}, \hat{w}+\kappa(\hat{p}))  \nonumber \\ 
    & + (x-1) \bigg[\hat{w}_{xx} + \kappa''(\hat{p}) \hat{D}^2  f^2(\hat{p}, \hat{w}+\kappa(\hat{p})) \nonumber \\ 
    &+\hat{D}  \kappa'(\hat{p}) \frac{df}{dx} (\hat{p}, \hat{w}+\kappa(\hat{p}))\bigg] \nonumber \\
    &-\hat{D}\kappa'(\hat{p}) G(x, t) \nonumber \\ &- \hat{D}^2 \bigg[\frac{\partial f}{\partial \hat{p}}(\hat{p}, \hat{w}+\kappa(\hat{p}))\kappa'(\hat{p}(x, t)) \nonumber \\ & + \kappa''(\hat{p})f(\hat{p}, \hat{w}+\kappa(\hat{p})) \bigg] \int_0^x \hat{\Phi}(x, y, t) G(y, t) dy \label{eq:g7-def}\\ 
    g_8 =& g_{4x} \nonumber \\ 
    &= \hat{D}^2 \bigg[f(\hat{p}, \hat{w}+\kappa(\hat{p}))^\top \kappa''(\hat{p}) \nonumber \\ &+ \kappa'(\hat{p})\frac{\partial f}{\partial \hat{p}}(\hat{p}, \hat{w}+\kappa(\hat{p})) \bigg]\hat{\Phi}(x, 0, t) \label{eq:g8-def}
\end{align}
\begin{align}
     g_9 =& \hat{w}_{xx}+  f(\hat{p}, \hat{w}+\kappa(\hat{p}))^\top \kappa''(\hat{p}) \hat{D}^2 \nonumber \\ & \times 
     f(\hat{p}, \hat{w}+\kappa(\hat{p})) \nonumber \\ & + \kappa'(\hat{p})\hat{D} \frac{d}{dx}[f(\hat{p}, \hat{w}+\kappa(\hat{p}))]  \nonumber \\ 
    & +  \bigg[\hat{w}_{xx} + f(\hat{p}, \hat{w}+\kappa(\hat{p}))^\top \kappa''(\hat{p})\hat{p}_x \hat{D}   \nonumber \\ & +\hat{D}  \kappa'(\hat{p}) f_x(\hat{p}, \hat{w}+\kappa(\hat{p}))\bigg] \nonumber \\
      & + (x-1) \bigg[\hat{w}_{xxx}\frac{d}{dx} \bigg[ f(\hat{p}, \hat{w}+\kappa(\hat{p}))^T\kappa''(\hat{p})\hat{D}^2 \nonumber \\ &  \times   f(\hat{p}, \hat{w}+\kappa(\hat{p})) \nonumber \\ &+\hat{D}  \kappa'(\hat{p}) \frac{df}{dx}(\hat{p}, \hat{w}+\kappa(\hat{p}))\bigg]\bigg] -\frac{d}{dx} \bigg[\hat{D}\kappa'(\hat{p}) G(x)  \nonumber \\ 
      & -\hat{D}\kappa'(\hat{p}(x, t)) \int_0^x \hat{D}f(\hat{p}(x, t), \hat{w}(x, t)+\kappa(\hat{p}(x, t)))\nonumber \\ & \times \hat{\Phi}(x, y, t) G(y,t) dy\nonumber \\ 
    & -\hat{D} \kappa''(\hat{p})\hat{p}_x \int_0^x \hat{\Phi}(x, y, t) G(y) dy  \bigg]\label{eq:g9-def} \\ 
    g_{10} =& g_{8x} \nonumber \\ 
    =& \hat{D}^2 \frac{d}{dx}\bigg[f(\hat{p}, \hat{w}+\kappa(\hat{p}))^\top \kappa''(\hat{p}) \nonumber \\ & + \kappa'(\hat{p})\frac{\partial f}{\partial \hat{p}}(\hat{p}, \hat{w}+\kappa(\hat{p})) \bigg]\nonumber \\ & + \hat{D}^3 \frac{\partial f}{\partial \hat{p}}(\hat{p}, \hat{w}+\kappa(\hat{p})) \hat{\Phi}(x, 0, t)  \nonumber \\ & \times \bigg[f(\hat{p}, \hat{w}+\kappa(\hat{p}))^\top \kappa''(\hat{p}) \nonumber \\ &  + \kappa'(\hat{p})\frac{\partial f}{\partial \hat{p}}(\hat{p}, \hat{w}+\kappa(\hat{p})) \bigg]
\end{align}
where
\begin{align}
    G(y, t) :=& f(\hat{p}(y, t), \hat{w}(y, t) + \kappa(\hat{p}(y, t))) \nonumber  \\ 
    & + \frac{\partial f(\hat{p}(y, t), \hat{w}(y, t) + \kappa(\hat{p}(y, t))}{\partial \hat{u}}(y-1) \nonumber \\ 
    & \times \bigg(\hat{w}_x(y, t) + \kappa'(\hat{p}(y, t))\hat{D}(t) f(\hat{p}(y, t), \hat{w}(y, t) \nonumber \\ &+ \kappa(\hat{p}(y, t))) \bigg)
\end{align}

\section{Proof of technical Lemma's used in Section \ref{sec:stability} } \label{appendix:proof-tech-expressions}

\begin{lemma} \label{lemma:p-bound}
There exist class \(\mathcal{K}_\infty\) functions \(\alpha_{12}, \alpha_{13}\) such that, for all \(x \in [0,1]\),
\begin{align}
    |\hat{p}(x, t)| &\leq \alpha_{12}\big(|X| + \|\hat{u}[t]\|_{L^1}\big) \,, \label{eq:lemma1-bnd-1} \\ 
    |\hat{p}(x, t)| &\leq \alpha_{13}\big(|X| + \|\hat{w}[t]\|_{L^1}\big) \,. \label{eq:lemma1-bnd-2}
\end{align}
\end{lemma}

\begin{pf}
By definition, the predictor \(\hat{p}\) satisfies
\begin{align}
    \frac{\partial}{\partial x} \hat{p}(x, t) &= \hat{D}(t) f\big(\hat{p}(x, t), \hat{u}(x, t)\big), \\
    \hat{p}(0, t) &= X(t).
\end{align}
Since \(\hat{D}(t)\) is bounded in \([\underline{D}, \overline{D}]\) and the plant is strictly forward complete, there exists \(\alpha_{12} \in \mathcal{K}_\infty\) such that \eqref{eq:lemma1-bnd-1} holds.

Similarly, by the exact backstepping transform,
\begin{align}
    \frac{\partial}{\partial x} \hat{p}(x, t) &= \hat{D}(t) f\big(\hat{p}(x, t), \kappa(\hat{p}(x, t) + \hat{w}(x, t))\big), \\
    \hat{p}(0, t) &= X(t).
\end{align}
Applying the same reasoning yields \(\alpha_{13} \in \mathcal{K}_\infty\) such that \eqref{eq:lemma1-bnd-2} holds.
\end{pf}

\begin{lemma} \label{lemma:u-bound}
    There exists $\mathcal{K}_\infty$ function $\alpha_{14}, \alpha_{15}$ such that 
    \begin{align}
            |\hat{u}(x, t)| \leq&  \alpha_{14}(|X| + \|\hat{w}[t]\|_{L^1} + \|\hat{w}_x[t]\|_{L^1} ) + M_3\epsilon\,,  \label{eq:hatu-bound} \\
            |\tilde{u}(x, t)|\leq& \alpha_{15}(\|\tilde{u}_x(x, t)\|_{L^1})\,. \label{eq:tilde-bound}
    \end{align}
\end{lemma}
\begin{pf}
Applying integration by parts, substitution, the error bound of Theorem \ref{thm:uat-predictor}, Assumption \ref{assumption:growth-condition}, Lemma \ref{lemma:p-bound} and triangle inequality yields
\begin{align}
    |\hat{u}(x, t)| \leq& |\hat{u}(1, t)| + \int_0^1 |\hat{u}_x(x, t)|dx \nonumber \\ 
    \leq& \kappa(\hat{p}_{NO}(1, t)) + \int_0^1 |\hat{w}_x(x, t) + \kappa'(\hat{p})\hat{p}_x(x, t)| dx \nonumber \\  
    \leq& M_3 (\alpha_{13}(|X| + \|\hat{w}[t]\|_{L^1}) + \epsilon) + \|\hat{w}[t]\|_{L^1} \nonumber \\ & +M_4|\hat{D}(t)|[2M_1\alpha_{13}(|X| + \|\hat{w}[t]\|_{L^1}) \nonumber \\ & + \|\hat{w}[t]\|_{L^1}]
\end{align}
Thus, there exists $\alpha_{14}\in \mathcal{K}_\infty$ such that \eqref{eq:hatu-bound} holds. 
\eqref{eq:tilde-bound} is trivial given that $\tilde{u}(1, t) = 0$ and applying integration by parts yields $|\tilde{u}(x, t) | \leq \|\tilde{u}_x[t]\|_{L^1}$.
\end{pf}

\begin{lemma}\label{lemma:g1-g2}
    There exists $\alpha_{16}, \alpha_{17}, \alpha_{18} \in \mathcal{K}_\infty$ and constant $M_5 = \frac{1}{\underline{D}}$ such that
    \begin{align}
        \int_0^1 (1+x)& |g_1(x, t) |dx \nonumber \\ \leq& \alpha_{16}(|X(t)| + \|\hat{w}[t]\|_{L^1} + \|\hat{w}_x[t]\|_{L^1})\,,  \label{eq:g1-bound} \\ 
               \int_0^1 (1+x)& |g_2(x, t) |dx \nonumber \\ \leq& \alpha_{17}(|X(t)| + \|\hat{w}[t]\|_{L^1} + \|\hat{w}_x[t]\|_{L^1}) \label{eq:g2-bound} \,, \\ 
               |\tilde{u}_x(1, t)| \leq& |\tilde{D}(t)| \big[M_5|\hat{w}_x(1, t)| \nonumber \\ & +M_4 M_1M_3 \epsilon + \alpha_{18}(|X| + \|\hat{w}[t]\|_{L^1}) \big] \label{eq:utildex-1-bound}\,. 
    \end{align}
\end{lemma} 
\begin{pf}
By \eqref{eq:g1-def}, Assumption \ref{assumption:growth-condition}, Lemma \ref{lemma:p-bound}, and the triangle inequality we have 
\begin{align}
    \int_0^1 &(1+x)|g_1(x, t)| dx\nonumber \\  \leq& \frac{1}{\underline{D}} \int_0^1 (1+x)| \hat{w}_x(x, t)|dx\nonumber \\ & + \int_0^1 (1+x)| M_4 (M_1(|\hat{p}(x, t)|  + |\hat{w}(x,t)| \nonumber \\ & + |\kappa(\hat{p}(x, t))|)) dx  \nonumber  \\
    \leq &\frac{2}{\underline{D}} \|\hat{w}_x[t]\|_{L^1}+ 2M_4M_1 \|\hat{w}[t]\|_{L^1} \nonumber \\ &+ 2 M_4 M_1(1+M_3) \alpha_{13}(|X| +\|\hat{w}[t]\|_{L^1})\,. 
\end{align}
Hence $\alpha_{16}$ exists. In the same exact manner, one can show the existence of $\alpha_{17}$. For the bound on $|\tilde{u}_x(1, t)|$, we applied the Definition in \eqref{eq:uxt-2}, the triangle inequality, Assumption \ref{assumption:growth-condition}, Lemma \ref{lemma:p-bound}, and the bound on $\hat{D}(t)$. 
\end{pf}
\begin{lemma}\label{lemma:g3-g4}
    There exists $\alpha_{19:22} \in \mathcal{K}_\infty$ such that
    \begin{align}
        \int_0^1 &(1+x)|g_3(x, t)| dx \nonumber \\ \leq& \alpha_{19}(|X(t)| + \|\hat{w}[t]\|_{L^1} + \|\hat{w}_x[t]\|_{L^1} + \epsilon)\,, \label{eq:g3-bound} \\
        |f_{\tilde{u}}(t)| \int_0^1& (1+x)|g_4(x, t)| dx \nonumber \\  \leq& |\tilde{u}(0, t)|\times \alpha_{20}(|X(t)| + \|\tilde{u}_x[t]\|_{L^1} + \|\hat{w}[t]\|_{L^1} \nonumber \\ & + \|\hat{w}_x[t]\|_{L^1} + \epsilon)  \,, \label{eq:g4-bound} \\
        |\hat{w}_x(1, t)|& \nonumber \\ \leq& |\dot{\hat{D}}(t)| \alpha_{21}(|X(t)| + \|\hat{w}[t]\|_{L^1} + \|\hat{w}_x[t]\|_{L^1} + \epsilon) \nonumber\\ 
        &+\tilde{u}(0, t)| \times \alpha_{22}(|X(t)| + \|\hat{w}[t]\|_{L^1} \nonumber \\ & + \|\hat{w}_x[t]\|_{L^1} + \epsilon ) \,. \label{eq:whatx-1-bound} 
    \end{align}
\end{lemma}
\begin{pf}
    Notice that the transition matrix $\hat{\Phi}$ satisfies the bound
    \begin{align}
        |\hat{\Phi}(x, y, t)| \leq& \exp\bigg (\overline{D} \max_{x\in [0, 1]} \left|\frac{\partial f}{\partial \hat{p}}(\hat{p}(x, t), \hat{u}(x, t))\right|\bigg) \nonumber \\ \leq& \exp(\overline{D}M_2)\,, \label{eq:phi-uniform-bound}
    \end{align}
    as it satisifes the linear ODE for $y\in [x, 1]$
    \begin{align}
       \frac{\partial \hat{\Phi}(x, y, t)}{\partial t} =& \hat{D}(t) \frac{\partial f}{\partial \hat{p}}(\hat{p}(x, t), \hat{u}(x, t)) \hat{\Phi}(x,y, t) \,,\label{eq:dphi-dt} \\ 
       \hat{\Phi}(x, x, t) =& 1\,. 
    \end{align}
    Noting that $f$ is continuously differentiable, by the mean value theorem and Lemma \ref{lemma:u-bound}, we have that there exists a $\mathcal{K}_\infty$ function $\alpha_{48}$ such that 
    \begin{align}
       \bigg| &\frac{\partial f(\hat{p}(y, t), \hat{w}(y, t) + \kappa(\hat{p}(y, t))}{\partial \hat{u}} \bigg| \nonumber \\ & \hspace{10pt} \leq  \alpha_{48}(|X| + \|\hat{w}[t]\|_{L^1} + \|\hat{w}_x[t]\|_{L^1} + \epsilon)\,.\label{eq:hatu-deriv-bound}
    \end{align}
    Combining the bounds on $\hat{\Phi}$ as well as the bound on the partial derivative, one can follow a similar calculation as Lemma \ref{lemma:g1-g2} to achieve $\alpha_{19}$. 
    To obtain $\alpha_{20}$, notice that, by mean value theorem, there exists $d_2$ between $u(0, t)$ and $\hat{u}(0, t)$ such that 
    \begin{align}
        f_{\tilde{u}}(t) = \frac{\partial f}{\partial \hat{u}}(X(t), d_2(t)) \tilde{u}(0, t)\,. 
    \end{align}
    Then, we have that there is some $\alpha_{49}, \alpha_{50} \in \mathcal{K}_\infty$
    \begin{align}
        |f_{\tilde{u}}(t)| \leq & |\tilde{u}(0, t)| \times \alpha_{49}(|X(t)| + |u(0, t)| + |\hat{u}(0, t)|)  \nonumber \\ 
        \leq& |\tilde{u}(0, t)| \times  \alpha_{49}(|X(t)| + |\tilde{u}(0, t)| + 2|\hat{u}(0, t)|) \nonumber \\  
        \leq& |\tilde{u}(0, t)|\times  \alpha_{50}(|X(t)| + \|\tilde{u}_x\|_{L^1} + \|\hat{w}[t]\|_{L^1} \nonumber \\ & + \|\hat{w}_x[t]\|_{L^1} + \epsilon)  \label{eq:futilde-bound}
    \end{align}
    Applying the result, and noting that $g_4$ is easily bounded yields $\alpha_{20}$. 

    Lastly, to bound $|\hat{w}_x(1, t)|$, we can reapply the exact techniques above to obtain $\alpha_{21}$ and $\alpha_{22}$.  
\end{pf}
\begin{lemma} \label{lemma:g5-g6}
    There exists $\alpha_{23}, \alpha_{24}$ in $\mathcal{K}_\infty$ such that 
  \begin{align}
        \int_0^1& (1+x) |g_5(x, t) |dx \nonumber \\ \leq& \alpha_{23}(|X(t)| + \|\hat{w}[t]\|_{L^1} + \|\hat{w}_x[t]\|_{L^1} +  \|\hat{w}_{xx}[t]\|_{L^1} + \epsilon) \label{eq:g5-bound}\,,  \\ 
               \int_0^1& (1+x) |g_6(x, t) |dx \nonumber \\ \leq& \alpha_{24}(|X(t)| + \|\hat{w}[t]\|_{L^1} + \|\hat{w}_x[t]\|_{L^1} +  \|\hat{w}_{xx}[t]\|_{L^1} + \epsilon) \label{eq:g6-bound}\,.
    \end{align}
\end{lemma}
\begin{pf}
First, using the definition $g_5$, it is clear that there exists $\alpha_{51}, \alpha_{52}, \alpha_{53} \in \mathcal{K}_\infty$ such that 
    \begin{align}
     \int_0^1 &(1+x)| \hat{w}_{xx}(x,t)|dx \leq\alpha_{51}(\|\hat{w}_{xx}[t]\|_{L^1})\,, \label{eq:wxx-bound} \\
     \int_0^1 &(1+x)|\kappa'(\hat{p}(x, t))\hat{D}^2(t)  f^2(\hat{p}(x, t), \hat{w}(x, t)\nonumber \\ &+\kappa(\hat{p}(x, t)))|dx \nonumber \\  \leq& \alpha_{52} ( |X| + \|\hat{w}[t]\|_{L^1}) \,, \label{eq:kprime} \\
     \nonumber  \int_0^1 &(1+x)|\hat{D}(t)  \kappa(\hat{p}(x, t) \frac{d}{dx} f(\hat{p}(x, t), \hat{w}(x, t)\\ &+\kappa(\hat{p}(x, t)))|\nonumber \\ \leq& \alpha_{53}(|X| + \|\hat{w}[t]\|_{L^1} + \|\hat{w}_x[t]\|_{L^1} + \epsilon)\,, 
     \end{align}
     \eqref{eq:wxx-bound} is easily bounded by twice the norm. To show \eqref{eq:kprime}, we apply the standard bounds on $\hat{D}$ and the Lipschitz bound of $f$ with bounds on $\hat{p}$. To achieve the last bound, notice that we have a bound on $\frac{\partial f}{\partial \hat{u}}$ given in \eqref{eq:hatu-deriv-bound} and applying Assumption \ref{assumption:growth-condition} to obtain the bound on $\frac{\partial f}{\partial \hat{p}}$ yields the result. The bound of \eqref{eq:g6-bound} is a direct consequence of the bound on \eqref{eq:g5-bound} and \eqref{eq:g1-bound} with the projection of the update law.
\end{pf}

\begin{lemma} \label{lemma:g7-g8}
    There exists $\alpha_{25}, \alpha_{26}  \in \mathcal{K}_\infty$ such that
        \begin{align}
        \int_0^1 &(1+x)|g_7(x, t)| dx \nonumber \\ \leq& \alpha_{25}(|X(t)| + \|\hat{w}[t]\|_{L^1} + \|\hat{w}_x[t]\|_{L^1} + \|\hat{w}_{xx}[t]\|_{L^1} \nonumber \\ &+ \epsilon)\,, \label{eq:g7-bound} \\
        |f_{\tilde{u}}(t)| \int_0^1& (1+x)|g_{8}(x, t)| dx \nonumber \\ \leq& |\tilde{u}(0, t)| \times \alpha_{26}(|X(t)| + \|\hat{w}[t]\|_{L^1} + \|\hat{w}_x[t]\|_{L^1} \nonumber \\ &  +\|\tilde{u}_x[t]\|_{L^1}  +\epsilon ) \,. \label{eq:g8-bound}
    \end{align}
\end{lemma}
\begin{pf}
    We note that there exists $\alpha_{54:61} \in \mathcal{K}_\infty$ using the same calculation approaches as above:
    \begin{align}
        \int_0^1& (1+x)|\hat{w}_x(x, t)| dx \leq  \alpha_{54}(\|\hat{w}_x[t]\|_{L^1})\,, \label{eq:wx-obvious} \\
        \int_0^1 &(1+x)|\kappa'(\hat{p})\hat{D}(t) f(\hat{p}(x, t), \hat{w}(x, t)+\kappa(\hat{p}(x, t)))| dx \nonumber \\ \leq& \alpha_{55}(|X| + \|\hat{w}[t]\|_{L^1})\,,  \label{eq:kprime-obvious}\\
        \int_0^1 &(1+x)|(x-1)\hat{w}_{xx}(x, t)| dx \leq  \alpha_{56}(\|\hat{w}_{xx}[t]\|_{L^1})\,,  \label{eq:wxx-obvious} \\
        \int_0^1 &(1+x)|(x-1)\kappa''(\hat{p}(x, t))\hat{D}^2(t) \nonumber \\ & \times  f^2(\hat{p}(x, t), \hat{w}(x, t)+\kappa(\hat{p}(x, t)))| dx \nonumber \\ \leq& \alpha_{57}(|X| + \|\hat{w}[t]\|_{L^1})\,, \label{eq:kdoubleprime} \\
        \int_0^1 &(1+x)|(x-1)\hat{D}(t)  \kappa(\hat{p}(x, t)  \nonumber \\ & \times \frac{d}{dx} f(\hat{p}(x, t), \hat{w}(x, t)+\kappa(\hat{p}(x, t)))| dx \nonumber \\ \leq& \alpha_{58}(\|\hat{w}[t]\|_{L^1})\,, \label{eq:final-obvious} \\ 
        \int_0^1 &(1+x)|\hat{D}(t) \kappa'(\hat{p}(x, t)) G(x, t)| dx \nonumber\\ \leq& \alpha_{59}(|X| + \|\hat{w}[t]\|_{L^1} + \|\hat{w}_x[t]\|_{L^1} + \epsilon) \label{eq:gx}\,, \\
        \int_0^1 &(1+x)\bigg | \hat{D}(t)^2 \bigg[\frac{\partial f}{\partial \hat{p}}(\cdot, \cdot) \kappa'(\hat{p}(x, t)) \nonumber \\ & \times \int_0^x \hat{\Phi}(x, y, t)G(y)dy\bigg| dx \nonumber \\ \leq& \alpha_{60}(|X(t)| + \|\hat{w}[t]\|_{L^1} + \|\hat{w}_x[t]\|_{L^1} + \epsilon)\,, \label{eq:phi-1}
        \\ 
         \int_0^1 &(1+x)\bigg | \hat{D}(t)^2 \kappa''(\hat{p})f(\cdot, \cdot)
         ) \int_0^x \hat{\Phi}(x, y, t)G(y)dy\bigg| dx \nonumber \\ \leq& \alpha_{61}(|X(t)| + \|\hat{w}[t]\|_{L^1} + \|\hat{w}_x[t]\|_{L^1} + \epsilon)\,.\label{eq:phi-2}
    \end{align}
    \eqref{eq:wx-obvious} and \eqref{eq:wxx-obvious} are by definition. \eqref{eq:kprime-obvious} follows exactly as \eqref{eq:kprime}. To obtain a bound in \eqref{eq:kdoubleprime}, notice that via integration by parts we have $\int_0^1(1+x)|(x-1)\kappa''(\hat{p}(x, t))|dx \leq 3M_4$. \eqref{eq:final-obvious} is obtained in using the bound on $\kappa'$ and Assumption \ref{assumption:growth-condition}. \eqref{eq:gx} is obtained using \eqref{eq:hatu-deriv-bound}, Assumption \ref{assumption:growth-condition}, and the previous bounds. \eqref{eq:phi-1} is obtained using Assumption \ref{assumption:growth-condition} and the bound in \eqref{eq:phi-uniform-bound} . Lastly \eqref{eq:phi-2} is obtained using integration by parts, Assumption \ref{assumption:growth-condition} and the bound in \eqref{eq:phi-uniform-bound}.

    The bound of \eqref{eq:g7-bound} is a consequence of the sum of the bounds above and \eqref{eq:g8-bound} is a natural result given the bounds of \eqref{eq:kdoubleprime}, \eqref{eq:phi-uniform-bound}, and previously \eqref{eq:futilde-bound}.
\end{pf}

\begin{lemma}\label{lemma:g9-g10}
        There exists $\alpha_{27}, \alpha_{28} \in \mathcal{K}_\infty$ such that
        \begin{align}
        \int_0^1 &(1+x)|g_9(x, t)| dx \nonumber \\  \leq& |\hat{w}_{xx}(0, t)| + \alpha_{27}(|X(t)| + \|\hat{w}[t]\|_{L^1} + \|\hat{w}_x[t]\|_{L^1} \nonumber \\ & + \|\hat{w}_{xx}[t]\|_{L^1} + \epsilon)\,, \label{eq:g9-bound} \\
        |f_{\tilde{u}}(t)| \int_0^1 &(1+x)|g_{10}(x, t)| dx \nonumber \\ \leq&   |\tilde{u}(0, t)|  \times \alpha_{28}(|X(t)| + \|\hat{w}[t]\|_{L^1} + \|\hat{w}_x[t]\|_{L^1} \nonumber \\ & + \|\hat{w}_{xx}[t]\|_{L^1} +\|\tilde{u}_x[t]\|_{L^1} )  \,. \label{eq:g10-bound}
    \end{align}
\end{lemma}
\begin{pf}
    The proofs follow the exact same iteration as Lemma \ref{lemma:g7-g8} repeated once more except the term $\int_0^1 (x+1)|(x-1)\hat{w}_{xxx}(x, t)|dx$ appearing in \eqref{eq:g9-bound} is handled via integration by parts in the same manner as \cite[Eqn. 158]{delphine}. The term's containing $\int_0^1 (x+1)|\kappa'''(\hat{p}(x, t))|dx$ can similarly be bounded by applying integration by parts twice and noting that $\hat{p}_{xx}(x, t) = \frac{d}{dx} \hat{D}(t) f(\hat{p}(x, t), \hat{u}(x, t))$ is bounded by Assumption \ref{assumption:growth-condition}. The rest of the bounds are straightforward given the list of bounds in Lemma \ref{lemma:g7-g8}. 
\end{pf}

\begin{lemma}
    There exists $\alpha_{29:34} \in \mathcal{K}_\infty$ such that 
    \begin{align}
        |\hat{w}&_{xx}(1, t)|  \nonumber \\  \leq & \left( |\dot{\hat{D}}(t)| + |\dot{\hat{D}}(t)^2 | + |\dot{\hat{D}}(t)^3| \right) \times \alpha_{29}(|X(t)| + \|\hat{w}[t]\|_{L^1} \nonumber \\ & \hspace{1cm} + \|\tilde{u}_x[t]\|_{L^1} + \|\hat{w}_x[t]\|_{L^1}+  \|\hat{w}_{xx}[t]\|_{L^1} + \epsilon) \nonumber \\ 
        & 
        + |\tilde{u}_x(0, t)|  
         \alpha_{30}(|X(t)| + \|\hat{w}[t]\|_{L^1} + \|\tilde{u}_x[t]\|_{L^1} \nonumber \\ & \hspace{1cm} + \|\hat{w}_x[t]\|_{L^1} +\epsilon )  \nonumber \\ &
        +    |\hat{w}_x(0, t)|  \alpha_{31}(|X(t)| + \|\hat{w}[t]\|_{L^1} + \|\tilde{u}_x[t]\|_{L^1} \nonumber\\ &  + \|\hat{w}_x[t]\|_{L^1} + \epsilon  ) \nonumber  \\ 
        &\hspace{1cm}  +|\tilde{u}(0, t)|  \alpha_{32}(|X(t)| + \|\hat{w}[t]\|_{L^1} + \|\tilde{u}_x[t]\|_{L^1}  \nonumber \\ & + \|\hat{w}_x[t]\|_{L^1} +\epsilon ) \nonumber \\ 
        & + |\ddot{D}(t)| \alpha_{33}(|X(t)| + \|\hat{w}[t]\|_{L^1} + \|\tilde{u}_x[t]\|_{L^1} \nonumber \\ & \hspace{1cm}  + \|\hat{w}_x[t]\|_{L^1}+\epsilon) \nonumber \\ 
        & + |\tilde{D}(t)|\alpha_{34}(|X(t)| + \|\hat{w}[t]\|_{L^1} + \|\tilde{u}_x[t]\|_{L^1}  \nonumber \\ & \hspace{1cm} + \|\hat{w}_x[t]\|_{L^1}+\epsilon)\,.
    \end{align}
\end{lemma}

\begin{pf}
    First, notice that $\hat{w}_{xt}(1, t)$ can be obtained by taking the time derivative of the boundary condition of $\hat{w}_x(1, t)$. Taking such a time derivative is a lengthy calculation involving the triangle inequality. One follows the same exact calculation as in \cite{breschpietri2013backsteppingtransformationinputdelay} replacing Lemma 7 when necessary with the corresponding $\epsilon$ perturbations. 
\end{pf}

\end{document}